\documentclass[aip,jcp,reprint,floatfix]{revtex4-1}

\pdfoutput=1

\usepackage[T1]{fontenc}
\usepackage[utf8x]{inputenc}

\usepackage[pdftex, bookmarks, pdffitwindow=false, pdfstartview=FitH, pdfdisplaydoctitle, colorlinks, plainpages=false, pdftitle={Free energy landscapes for homogeneous nucleation of ice for a monatomic water model}, pdfpagelabels, hypertexnames, pdflang={en}, hyperfootnotes=false, breaklinks]{hyperref}

\usepackage{graphicx}
\usepackage{amsmath,amssymb}

\usepackage{geometry}
\usepackage[obeyall]{siunitx}
\newunit{\barunit}{bar}
\newunit{\calorie}{cal}
\def\avg#1{\ensuremath{\left\langle #1 \right\rangle}}

\usepackage{fixmath}

\usepackage{mathptmx}
\usepackage[Symbol]{upgreek}

\begin{document}

\title{Free energy landscapes for homogeneous nucleation of ice for a monatomic water model}

\author{Aleks Reinhardt}
\author{Jonathan P. K. Doye}
\email{jonathan.doye@chem.ox.ac.uk}
\affiliation{Physical and Theoretical Chemistry Laboratory, Department of Chemistry, University of Oxford, Oxford OX1 3QZ, United Kingdom}
\date{21 December 2011}

\begin{abstract}
We simulate the homogeneous nucleation of ice from supercooled liquid water at \SI{220}{\kelvin} in the isobaric-isothermal ensemble using the MW monatomic water potential. Monte Carlo simulations using umbrella sampling are performed in order to determine the nucleation free energy barrier. We find the Gibbs energy profile to be relatively consistent with that predicted by classical nucleation theory;  the free energy barrier to nucleation was determined to be $\sim$18$\,k_\text{B}T$  and the critical nucleus comprised $\sim$85 ice particles. Growth from the supercooled liquid gives clusters that are predominantly cubic, whilst starting with a pre-formed subcritical nucleus of cubic or hexagonal ice results in the growth of predominantly that phase of ice only.
\end{abstract}



\maketitle 

\section{Introduction}

It is not unreasonable to expect water to freeze into ice when it is cooled below its freezing point of \SI{0}{\celsius} in normal conditions. However, despite everyday experience, it is possible to cool very pure water droplets down to approximately \SI{-45}{\celsius} at standard pressure without observing freezing.\cite{Jeffery1997, Pruppacher1997, Oxtoby1998} In the absence of an external nucleation seed, water can remain in its supercooled state for exceedingly long periods of time. Classical nucleation theory (CNT) predicts the existence of a kinetic barrier to nucleation due to a competition between the unfavourable interfacial free energy and the favourable bulk free energy of growing clusters, rendering homogeneous nucleation extremely slow, as a cluster of some critical size must spontaneously form before large-scale crystallisation can occur.\cite{Laaksonen1995, Auer2005, Anwar2011} In other words, homogeneous nucleation is a rare event, where the average time between nucleation events is many orders of magnitude greater than the duration of the event itself.

Although in practice, heterogeneous nucleation of ice dominates the freezing process, homogeneous nucleation is interesting not only as a starting point towards more practical results, but also because it is thought to be important in processes such as the formation of ice in clouds, governing both global climate change and short-term weather,\cite{Oxtoby1992, Baker1997, Benz2005, Hegg2009, Spichtinger2010} and the cryopreservation of biological tissues.\cite{Toner1990} On the other hand, the fact that homogeneous nucleation is a very rare event (or, equivalently, that supercooled water is metastable with respect to ice in the absence of external nucleants) is likewise crucial in some phenomena, where it is important that water remain a (supercooled) liquid, such as electrification and rainfall in clouds.\cite{Hagen1981, Pruppacher1995, Koop2000, Debenedetti2003}

As nucleation is a rare event, it is intrinsically difficult to study. There are two principal factors potentially governing the slowness of nucleation: the nucleation free energy barrier could be very large or the dynamics of the process, excluding the kinetic barrier to nucleation, might be inherently slow. Indeed, the highly ordered hydrogen-bonded liquid structure, especially at lower temperatures, slows diffusion considerably; so much so, in fact, that it has been suggested that the Gibbs energy of diffusion activation corresponding to the transfer of water molecules across the ice-water interface may be rate-determining.\cite{Pruppacher1995, Pruppacher1997, Kabath2006} However, other experimental work suggests that the temperature dependence of the nucleation rate primarily reflects the behaviour of the free energy barrier of nucleation rather than diffusion activation.\cite{Kraemer1999}

At atmospheric pressure below the freezing point, the thermodynamic form of water is hexagonal ice (ice I$_\text{h}$) with the lonsdaleite structure; an important metastable phase in these conditions is cubic ice (ice I$_\text{c}$) with the diamond structure in the oxygen atom arrangement. Both are proton disordered and obey the Bernal-Fowler ice rules.\cite{Bernal1933,Pauling1935,Petrenko1999} Cubic ice appears to form as the predominant (if metastable) phase between \SI{130}{\kelvin} and \SI{150}{\kelvin},\cite{Debenedetti2003} but is converted into hexagonal ice above approximately \SI{200}{\kelvin}. It is thought to play an important r\^{o}le in certain atmospheric processes.\cite{Murray2005,Peter2006,Shilling2006,Heymsfield2011} Although cubic ice is always the metastable form, the energy released on conversion into hexagonal ice is only between 13 and \SI{50}{\joule\per\mole}.\cite{Petrenko1999} By contrast, for the TIP4P model of water,\cite{Jorgensen1983} cubic ice may, depending on the precise conditions of the simulation, be more stable than hexagonal ice,\cite{Fennell2005} although the energy difference may be less than the measurement uncertainty.\cite{Vega2008} Both forms of ice have been observed in nucleation experiments in various conditions.\cite{Bartell1994, Kraemer1999, Murray2005, Murray2006}

Ice crystallisation has been studied in a significant number of simulations using all-atom models of water;\cite{Svishchev1994,Svishchev1996,Yamada2002,Matsumoto2002,Nada2003,Radhakrishnan2003b,Radhakrishnan2003,Fernandez2006,Vrbka2006, Vrbka2007, Carignano2007, Quigley2008,Brukhno2008, Pluharova2010,Pereyra2011,Weiss2011,Rozmanov2011}
however, despite the obvious importance of the process of ice nucleation, there have not been many successful attempts at simulating the process so far. Crystallisation has been observed in an individual molecular dynamics (MD) trajectory at a fixed density lower than the liquid state density,\cite{Matsumoto2002} and it remains the only successful brute-force simulation of ice nucleation with an all-atom model. Nucleation under `special' conditions has been observed, such as in simulations of very small numbers of particles in a strong electric field\cite{Svishchev1994,Svishchev1996} or near a surface;\cite{Vrbka2006, Pluharova2010} however, in these studies, the ice clusters that grow quickly span the simulation box. An alternative approach is to use rare event methods to compute the free energy landscapes for crystallisation.

Such free energy landscapes have recently been calculated in several studies. Radhakrishnan and Trout used two-dimensional umbrella sampling,\cite{Radhakrishnan2003b,Radhakrishnan2003} Quigley and Rodger used a metadynamics approach,\cite{Quigley2008} and Brukhno and co-workers used replica exchange umbrella sampling.\cite{Brukhno2008} The first two studies mentioned use `global' rather than `local' order parameters. Although these order parameters detect the increase in order as crystallisation occurs, they do not specifically refer to the formation of a local crystal nucleus, and so it is not totally clear to what the variables in the energy landscape physically refer.  Furthermore, there are grounds to think that the pathways that these order parameters enhance may not be representative of real ice nucleation.  An increase in these order parameters is dependent on the accumulation of local Steinhardt vectors\cite{Steinhardt1983} such that they do not cancel out. Once a sufficient random fluctuation occurs in one of the orientations, it is driven onwards, which ensures that all the ice-like molecules are orientated in a correlated fashion; indeed, the orientational coherence affects even those water molecules that remain liquid-like. This appears to be at odds with a local picture of nucleation, since a molecule's behaviour is not just influenced by its local environment, but also, in a non-physical way, by the orientation of (ice) molecules distant from it. The use of global order parameters to drive crystallisation has also been shown to induce entropic break-up of small clusters.\cite{TenWolde1996}

The study by Brukhno and co-workers used novel local order parameters based on a maximum projection director approach,\cite{Brukhno2008} which do allow the calculation of a free energy profile as a function of an order parameter measuring the increasing crystallinity of a nucleus. However, because their order parameter is not rotationally invariant, but induces crystallisation with a specific orientation with respect to the simulation box, this bias is likely to lead to a significant error in the free energy barrier, because it excludes many other possible pathways (which could have a lower free energy). The space-fixed nature of the order parameter also induces a non-local and non-physical orientational coherence in the growing ice cluster.

The ice form generated in these simulations has varied from pure cubic ice,\cite{Quigley2008} to pure hexagonal ice,\cite{Radhakrishnan2003b} and to mixtures with predominantly hexagonal ice.\cite{Brukhno2008} The free energy barriers computed in these studies are intriguing. All these simulations used the TIP4P water model and presented results at a temperature of \SI{180}{\kelvin} and at standard pressure,\cite{Note1} which represents an approximately \SI{22}{\percent} supercooling. Brukhno and co-workers find a free energy barrier of approximately $130\,k_\mathrm{B}T$, although this value should be largely discounted because of their use of a non-rotationally-invariant order parameter, and, as they themselves note, because their system was not fully equilibrated.

The barriers reported by Radhakrishnan and Trout ($63\,k_\mathrm{B}T$) and Quigley and Rodger ($79\,k_\mathrm{B}T$)\cite{Note2} are more interesting. If the equilibrium free energy landscapes have been obtained, these values should be lower bounds to the free energy barrier one would get for the `perfect' order parameter because of configurations at the top of the barrier that are not truly representative of the intermediates between the two states. As one would generally expect an order parameter that follows the largest crystalline cluster within the system to provide a better representation of the intermediates, one would expect the free energy barrier obtained with such an approach to be larger than that obtained using global orientational order parameters. Classical nucleation theory can provide an estimate of the former; the CNT expression for the Gibbs energy barrier of a spherical cluster comprising $N$ particles is $\upDelta G (N) = -N \upDelta_\text{fus}\mu  + \gamma [ 36\uppi  ( N/\rho)^2  ]^{1/3}$,
where $\rho$ is the number density of the crystalline phase, $\upDelta_\text{fus} \mu$ is the change in chemical potential on fusion and $\gamma$ is the interfacial free energy between the two phases.\cite{Anwar2011}  An estimation of the classical nucleation theory result for TIP4P water at \SI{180}{\kelvin} gives a free energy barrier of $\upDelta G \approx 35\,k_\text{B}T$ and a critical nucleus size of 106. To obtain these, we use the simple approximation\cite{Oxtoby1992} that $\upDelta_{\text{fus}} \mu(T) \approx \upDelta_{\text{fus}} H_\mathrm{m} ((T_\text{fus}-T)/T_\text{fus})$, and we insert the appropriate values for the TIP4P model.\cite{Note3}\nocite{Vega2005,Handel2008} 
We note that one generally expects the interfacial free energy to decrease with decreasing temperature, as has been inferred from experiment,\cite{Bigg1953,Wood1970} and so the use of $\gamma$ from coexistence likely leads to an overestimation of this barrier.

One is tempted to ask why there is such a big difference in the barrier, and in the opposite direction to what is expected. The first option is that classical nucleation theory's shortcomings are responsible for a severe underestimation of the free energy barrier; however, although it is well-known that there are deficiencies in the framework of CNT, such a big difference nevertheless seems surprising. Another option might be that the free energy landscapes computed in simulations are not fully at equilibrium; to find the free energy barrier of nucleation, not only the end points, but the entire path space must be equilibrated. The latter may have been a problem in the previous simulations, as the nature of the order parameters may bias the system to locate more easily pathways that are high in free energy, but where the dynamics along the reaction co-ordinate (once the free energy barrier has been negated) are more computationally tractable to observe. For example, the coherence induced by the use of global order parameters may make it easier for water molecules to join the ice crystal if they can `feel' the orientation of the crystal, even if this is not the natural mode of growth.

Given the importance of water, it is perhaps surprising that free energy profiles for ice nucleation using a local, rotationally invariant order parameter have not been computed for an all-atom model of water. The problem seems to be that the natural dynamics associated with ice crystallisation are extremely slow. For example, even the growth by a few layers of a planar ice-water interface below the freezing point, which is a purely downhill process in free energy, can take weeks of computer time with a TIP4P-type model.\cite{Fernandez2006} The fastest rate of growth of such an interface is just below the freezing point,\cite{Pereyra2011,Weiss2011,Rozmanov2011} but this is where the free energy barrier to nucleation is very large. Intriguingly, it has been shown that ice crystal growth occurs more rapidly in molecular dynamics simulations than in corresponding Monte Carlo ones,\cite{Fernandez2006} which is not typically the case in simulations of other materials. One possible rationalisation for this is that collective motion possible in molecular dynamics simulations helps to speed up the dynamics of cluster reorganisation in the process of crystallisation.

Here, we present simulations with a somewhat more limited aim, but which we hope present a step towards understanding the homogeneous nucleation of ice. We wish to look at a model of water that does not have the full complexity of TIP4P-type models, namely the monatomic model of water (MW) proposed by Moore and Molinero,\cite{Molinero2009} which has already been used to study both the supercooled liquid\cite{Moore2009,Limmer2011} and ice crystallisation (in the bulk\cite{Moore2010, Moore2011, Li2011} and in confinement\cite{Kastelowitz2010, Moore2010b}). Furthermore, some nucleation rates and critical cluster sizes have very recently been reported from molecular dynamics simulations near the liquid transformation temperature of \SI{202}{\kelvin}\cite{Moore2011} and in the range \SIrange{220}{240}{\kelvin}.\cite{Li2011}
This potential is essentially a modification of the three-body Stillinger-Weber potential\cite{Stillinger1985} for silicon with a greater weight given to tetrahedrality. Despite being monatomic, it provides a surprisingly good structural and thermodynamic representation of water; however, unlike all-atom water models, it crystallises relatively readily.\cite{Moore2010} The absence of explicit hydrogens appears to change dramatically the rate of nucleation. This could be because of a change in the free energy barrier to nucleation, or because the dynamics of the process are much quicker. Indeed, it is the unnaturally fast diffusion in the model which is most at odds with experiment.\cite{Moore2010}  A possible rationalisation of this behaviour may be that the complication of hydrogen bond flipping is no longer an issue. For example, if one molecule were to flip in TIP4P water, to continue obeying the ice rules, a whole series of water molecules would need to flip, which would clearly be an unlikely event. When hydrogens are removed, this complication no longer exists. In this work, we make use of the model's enhanced crystallisability to compute the free energy profile for nucleation in the MW model and hence to understand better the thermodynamics of ice nucleation.

\section{Simulation methods}
\label{sect:methods}
To simulate the system, we use the standard Metropolis Monte Carlo approach. In addition, we use the umbrella sampling\cite{Torrie1977, Chandler1987} method with windowing in the isobaric-isothermal ensemble to calculate the free energy profile. In the umbrella sampling simulations, we iteratively update the weights depending on the number of steps spent at each value of the order parameter in the previous iteration. To test if the simulation is equilibrated, we need to ensure (a) that there is (approximately) even sampling throughout the window, and (b) that there is sufficient exchange between order parameter values. A further check to ensure that we have sufficiently equilibrated each window is to ascertain that the potential enthalpy as a function of the order parameter is consistent in the overlapping regions.

In simulations, care must be taken to ensure that the ice clusters do not begin to span the box.
Such configurations are a non-physical consequence of the periodic boundary conditions and should be excluded from our sampling of the nucleation free energy profile, but can arise due to fluctuations to a more cylindrical shape. To reduce the probability of spanning clusters arising, we simulate the nucleation process in a cubic box, as if we allow box dimensions to vary independently, random fluctuations in the liquid phase often produce long cuboid boxes in which the formation of spanning clusters is more likely and which can thus crystallise readily, albeit in a non-physical manner.
We should note that the spanning of clusters only happens very occasionally in our simulations because we use a sufficiently large number of particles that for the size of the crystalline clusters we consider, their surface to volume ratio is minimised by a spherical rather than a cylindrical shape.

\begin{figure}[tb]
\begin{center}
\includegraphics{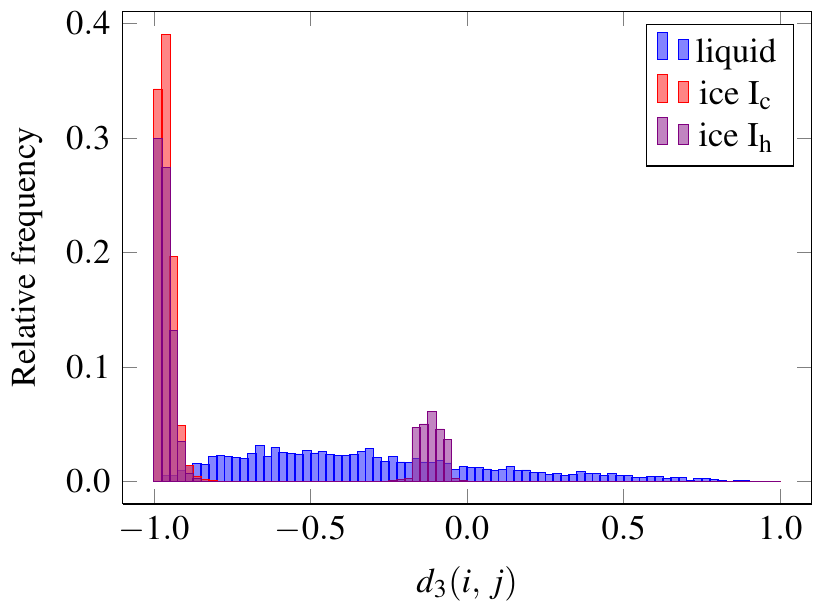}
\end{center}
\caption{A typical probability distribution for all pairs of $d_3(i,\,j) = \hat{\mathbold{q}}_3(i)\cdot \hat{\mathbold{q}}_3(j)$, where particles $i$ and $j$ are within \SI{3.6}{\angstrom} of each other. The three states depicted have all been equilibrated at \SI{220}{\kelvin} and the ice structures are not, therefore, `perfect'. This figure is analogous to those in Refs~\onlinecite{Moore2010b} and \onlinecite{Romano2011}; we reproduce it here for convenience.}\label{fig-MW-histogram-dotProds-3-all}
\end{figure}

The order parameter we use is the size of the largest ice cluster in the system. The choice of an ice classification parameter is not trivial, as it must on the one hand be local and be able to induce the growth of a small ice cluster, but on the other hand, it must be strict enough to encourage the growth of ice, rather than a structure only locally reminiscent of ice, but with no long-range order. We use Steinhardt-style local parameters\cite{Steinhardt1983, TenWolde1996} similar to those used previously in studies of tetrahedral liquids;\cite{Moore2010b, Romano2011} namely, we use the $(2l+1)$-dimensional real vector order parameter $\mathbold{q}_l(i)$ with components $m\in\left[-l,\,l\right]$ ($m\in\mathbb{Z}$) given by
\begin{equation}q_{lm}(i) =  \frac{1}{N_\text{neighs}(i)} \sum_{j=1}^{N_\text{neighs}(i)} S_{lm}\left(\theta_{ij},\,\varphi_{ij}\right),\end{equation}
where $S_{lm}$ are the spherical harmonics, $N_\text{neighs}$ is the number of neighbours of particle $i$,  $\theta$ and $\varphi$ are the polar angles, and we use $l=3$. For an up to five-fold decrease in computation time, we use real spherical harmonics\cite{Blanco1997} as opposed to complex ones. The sum in $j$ is over all neighbours within \SI{3.6}{\angstrom}. We then calculate $d_l(i,\,j) =  \hat{\mathbold{q}}_{l}(i) \cdot \hat{\mathbold{q}}_{l}(j)$. The resulting quantity ranges between $-1$ and $+1$, with cubic ice exhibiting a peak at $-1$ and hexagonal ice exhibiting a large peak at $-1$ (corresponding to staggered bonds) and a small peak near $-0.1$ (corresponding to eclipsed bonds). By plotting the distribution of $d_3$ values for the liquid and the solid phases (Fig.~\ref{fig-MW-histogram-dotProds-3-all}), a critical threshold can be determined as the first point where the probability of being in the solid phase is zero. This scheme works very well with fully-formed crystals and liquid water, which are thus excellently differentiated; however, if we used this as our classification parameter in attempting to grow ice clusters, this would unfairly bias us towards cubic ice, as we would accept imperfectly formed cubic ice (with 3 connections) as ice-like, but imperfectly formed hexagonal ice with three suitable connections would not be accepted as ice-like if two of these connections were in the $d_3\le -0.82$ region and one were in the vicinity of $-0.1$. If we wish to ensure that we do not bias simulations towards cubic ice, we need to be careful about our classification definitions.\cite{Romano2011} Our overall classification parameter is thus defined as
\begin{equation}n_\text{con}(i) = \sum_{j=1}^{N_\text{neighs}(i)}  \Gamma(d_3(i,\,j)),\end{equation}
where
\begin{equation}\Gamma(x) = \begin{cases} 1  \qquad & \text{if } [(x < -0.82) \wedge (-0.145 < x < -0.065)]  \\ 0 & \text{otherwise}. \end{cases}
\end{equation}
We classify a particle as ice-like if $n_\text{con} \ge 3$ and as liquid-like otherwise. This gives perfect identification in both equilibrated cubic and equilibrated hexagonal ice. Although it appears that liquid water, having a considerable proportion of particles in the $d_3$ region where ice I$_\text{h}$ has its second peak, might be mistakenly classified as ice, it is in fact almost nowhere that it has a sufficient number of connections with this value of $d_3$, and we find a misidentification rate in supercooled liquid water of approximately \SI{0.8}{\percent}, suggesting that this is an excellent classification parameter. The order parameter we use in umbrella sampling simulations to track nucleation is the size of the largest cluster of particles classified as ice, where two particles belong to the same crystalline cluster if they are both ice-like and within \SI{3.6}{\angstrom} of each other.

To calculate nucleation rates, we perform MD simulations using the \textsc{Lammps} molecular dynamics code\cite{Plimpton1995} in a Bennett-Chandler-type approach.\cite{Auer2001,Auer2002, Auer2005, Filion2010} We assume that the crystal nucleation rate $R$ is related to the Gibbs energy barrier $\upDelta G$ through an Arrhenius-type equation $R = \kappa P(n_\text{crit})$, where $n_\text{crit}$ is the size of the critical cluster, $P(n_\text{crit})=\exp(-\upDelta G(n_\text{crit})/k_\text{B}T)$ and $\kappa$ is the kinetic pre-factor. We can assume\cite{Auer2001} that the kinetic pre-factor can be written as $\kappa = Z\rho_\text{liq} f_{n_\text{crit}}$,
where $Z^2=\left|\upDelta G^{\prime\prime}(n_\text{crit})\right| / 2\uppi k_\text{B}T$ is the Zeldovich factor, $f_{n_\text{crit}}$ is the rate at which particles are attached to the critical cluster, and $\rho_\text{liq}$ is the number density of the supercooled liquid. The attachment rate $f_{n_\text{crit}}$ can be expressed as the diffusion of the cluster size at the top of the barrier\cite{Filion2010} by a modified Einstein relation, $f_{n_\text{crit}} = \lim_{t\to\infty} \avg{ (n(t)-n_\text{crit})^2}/2t$. Although this is considerably simpler than the full Bennett-Chandler scheme,\cite{Chandler1978,RuizMontero1997} it has been shown to agree very well with other methods of calculating rates in crystal nucleation, such as forward flux sampling and direct measurement from molecular dynamics.\cite{Filion2010}

\section{Results}
\subsection{Free energy profile}
\begin{figure}
\centering
\includegraphics{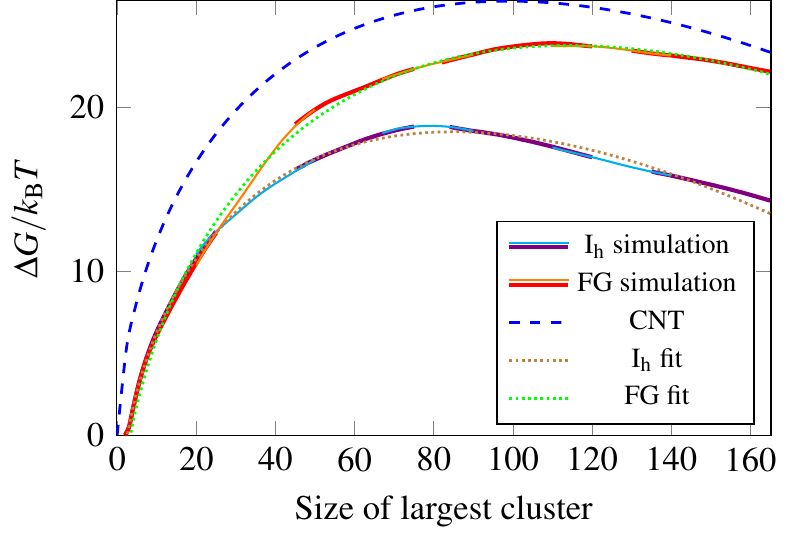}
\caption{The free energy profile of MW nucleation as a function of the size of the largest crystalline cluster in the system. Simulation results from different windows are depicted in alternating colours to show their overlap. The dashed line corresponds to the classical nucleation theory prediction; dotted lines depict fits to the simulation data. The free energy profiles for ice nucleation seeded with hexagonal crystal clusters (I$_\text{h}$) and ice nucleation directly from the supercooled liquid (FG) are shown.}\label{fig-MW-energybarrier}
\end{figure}

In our Monte Carlo simulations, we simulated 1400 MW particles at \SI{1}{\barunit} pressure and \SI{220}{\kelvin}, which corresponds to a \SI{20}{\percent} supercooling. The umbrella sampling used partially overlapping windows of various sizes until results were consistent and the criteria for equilibration given in Section~\ref{sect:methods} were deemed to have been fulfilled. On average, this involved approximately $5\times 10^{10}$ Monte Carlo steps for each window. The simulations involved three distinct scenarios: in one case, crystal clusters were grown directly from the supercooled liquid; in the other two cases, we started simulations in the initial few umbrella sampling windows with seed clusters of suitably equilibrated hexagonal and cubic ice, respectively. These were then allowed to shrink and grow, and clusters that grew into overlapping window regions were taken as initial starting points for the umbrella sampling simulations in later windows.  We must emphasise that the cubic and hexagonal ice simulations were not constrained in any way to form a particular phase of ice; we merely introduced small clusters taken from a particular crystal structure into the liquid, and the growth and shrinkage proceeded unconstrained from those starting configurations.\cite{Note4} The free energy profiles corresponding to crystal nucleus growth from the supercooled liquid and from equilibrated hexagonal clusters are shown in Fig.~\ref{fig-MW-energybarrier}; for clarity, we have not plotted the results for nucleation starting from cubic clusters, as the results are essentially identical to the hexagonal ice case.

\begin{figure*}
\centering
\includegraphics{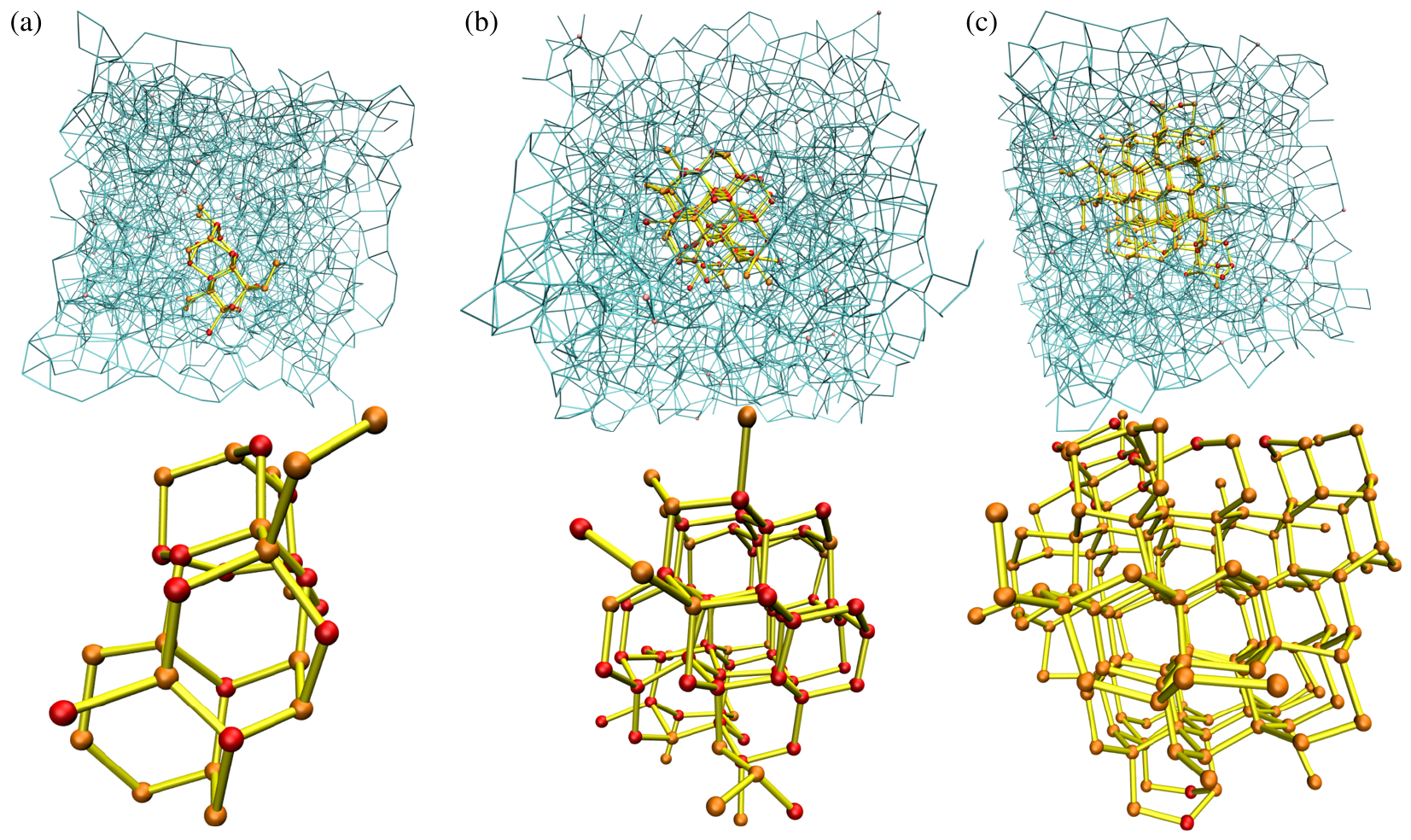}
\caption{Representative nucleation snapshots from umbrella sampling simulations. Particles within \SI{3.6}{\angstrom} are connected with lines; the lines are yellow within the largest ice cluster and cyan elsewhere (in the top row only). Particles are not connected across the periodic boundary. Spheres represent particles classified as ice: red spheres correspond to cubic ice, orange spheres correspond to hexagonal ice and pink spheres (in the top row only) correspond to ice particles not within the largest crystalline cluster.  In (a), a 30-particle cluster as nucleated from the supercooled liquid is shown. In (b), an 83-particle cluster as grown in a simulation initially seeded with an equilibrated cubic ice cluster is shown. In (c), a 165-particle cluster as grown in a simulation initially seeded with an equilibrated hexagonal ice cluster is shown. In each case, the top and bottom pictures depict the same cluster from different perspectives; one within the liquid framework and one showing solely the largest crystalline cluster.}\label{fig-MW-nucl-snapshots}
\end{figure*}

\begin{figure}
\centering
\includegraphics{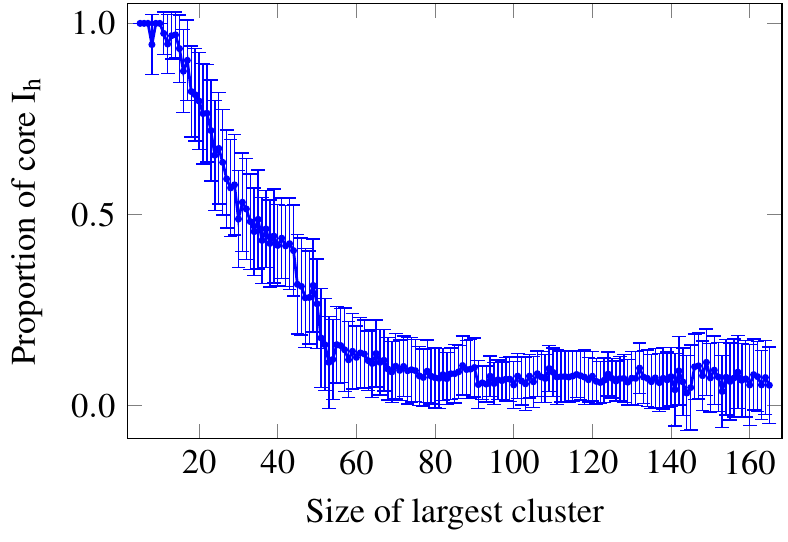}
\caption{The proportion of core ice particles classified as hexagonal for the set of simulations in which the crystalline cluster was grown directly from the supercooled liquid. Error bars show the standard deviation for the population of configurations at each ice cluster size.}\label{fig-MW-coreIh}
\end{figure}

Some snapshots of the nucleation process are shown in Fig.~\ref{fig-MW-nucl-snapshots}. Interestingly, the type of ice that grows during this nucleation process depends on the starting point. If we begin a simulation from the supercooled liquid, the ice clusters that form are initially mixed phases of hexagonal and cubic ice. As these `mixed' clusters grow, however, they become predominantly cubic (except for surface particles, for which a classification as cubic or hexagonal is less well-defined\cite{Note5}); the proportion of core\cite{Note6} ice particles  classified as being hexagonal drops from almost unity to approximately \SI{10}{\percent} by the time the crystal nucleus comprises 50 particles (Fig.~\ref{fig-MW-coreIh}). Starting from pre-formed cubic clusters in the lower umbrella sampling windows also results in primarily cubic ice growth. By contrast, starting the simulation with a pre-formed hexagonal cluster close in size to, but slightly smaller than, the critical size, results in essentially pure hexagonal ice growth. This suggests that in this model, ice prefers to grow in the phase of the underlying crystal nucleus. We have not been able to find a significant difference in the nucleation free energy profiles corresponding to cubic and hexagonal ice. However, our simulations show that the free energy barrier to nucleation of freely grown (mainly cubic) ice is $\sim$24$\,k_\text{B}T$ compared to $\sim$18$\,k_\text{B}T$ for pre-formed (hexagonal or cubic) ice, and the critical size is $\sim$114 compared to $\sim$85 (see Fig.~\ref{fig-MW-energybarrier}). The primary difference arises in the early stages of nucleation, where the ice grown directly from the supercooled liquid relaxes from mixed clusters to a much purer form of cubic ice. Nevertheless, the `cubic' ice grown from a supercooled liquid has more hexagonal defects than the corresponding cubic ice grown from a small perfectly cubic crystal nucleus.

In a very recent study\cite{Li2011} of homogeneous ice nucleation in the MW model, Li and co-workers studied the kinetics of nucleation using forward flux sampling at a range of temperatures, including at \SI{220}{\kelvin}. They found a critical crystal nucleus size of 265, which seems at first glance to be a considerably larger cluster than the sizes we obtain. However, the order parameter used in their study was different from the one we use; the principal difference is that all the neighbours of their classified ice particles were counted as being ice-like in their work (called `surface' particles), whilst we treat such particles as liquid-like. We calculated the sizes of our critical clusters using their order parameters (utilising $l=6$ spherical harmonics, a real-space neighbour cutoff of \SI{3.2}{\angstrom} and a dot product cutoff of $0.5$). When surface particles are not included, the cluster size is approximately equal to the values we obtained; however, including surface particles gives critical sizes of approximately 150 and 200 (corresponding to sizes of 85 and 114 without surface particles, respectively). Although this is still less than their reported 265, the disagreement is not nearly as significant as might initially appear. We can rationalise the discrepancy by noting that our system is able to relax locally due to the equilibration afforded it by umbrella sampling, whilst their system is driven forward aggressively by the forward flux sampling algorithm.

Similarly, Moore and Molinero have recently performed crystallisation simulations near the liquid transformation temperature;\cite{Moore2011} they estimate that the critical nucleus at \SI{208}{\kelvin} comprises 120 particles, whereas our simulations suggest, following histogram reweighting, that the critical nucleus at that temperature contains approximately 50 particles for the simulations grown directly from the supercooled liquid. As with the study by Li and co-workers, the discrepancy arises primarily from Moore and Molinero's inclusion of what they call `intermediate ice' in their largest cluster.

The question of whether `surface' particles are ice-like or liquid-like is not easy to answer, and whether restrictive or generous order parameters are more realistic is unclear; this is indeed a rather general problem affecting all nucleation studies.\cite{Lechner2011} It is certainly the case in our experience that particles near the surface of `core' ice can be very neatly arranged and thus reminiscent of the ice phase; but equally, there are many examples of particles adjacent to the ice cluster that have a structure totally incompatible with ice (such as having more than five neighbours). Nevertheless, although the choice of order parameter will affect the appearance of the free energy landscape (and the sampling efficiency), it should in principle have little effect on the barrier height and no effect on the rate of nucleation, provided that it is still a good parameter to describe the nucleation process.

Moore and Molinero studied the crystallisation of supercooled MW model water at \SI{180}{\kelvin}, where they observed ice formed from the supercooled liquid to consist of intercalated layers of cubic and hexagonal ice in the approximate ratio $2\,:\,1$.\cite{Moore2010,Moore2011b} We have not seen such behaviour in our simulations; an obvious possible reason for this difference is that their simulations were run at a lower temperature: indeed, they were run below the liquid transformation temperature of \SI{202}{\kelvin},\cite{Moore2011} where the structure of the liquid phase changes. However, Li and co-workers observed the ice grown in their simulations to be initially hexagonal, but transforming to a mixture of cubic and hexagonal ice in the approximate ratio $1\,:\,1$ in later stages of the nucleation at temperatures including \SI{220}{\kelvin}.\cite{Li2011} Umbrella sampling allows clusters both to grow and to shrink and thus to equilibrate locally, whereas their observations may be a result of kinetics dominating the growth pathway. In this regard, it is noteworthy that we have occasionally observed a hexagonal defect on cubic ice leading to further hexagonal ice growth and vice versa, and faster growth encourages more defects.

It is also interesting to note that the system can become trapped in different regions of path space. This is the case not only for `pure' cubic and hexagonal ice clusters, for which we would expect it to be difficult to interconvert once growth has started, but also between the ice clusters grown directly from the supercooled liquid, which were predominantly cubic, and the lower free energy pure cubic ice clusters. Such non-equilibrium effects are not new in studies of nucleation; for example, analogous behaviour has been observed in the crystal nucleation of a binary suspension of oppositely charged colloids.\cite{Sanz2007}

From classical nucleation theory, we can estimate the free energy of nucleation using the formula given in the Introduction.  We can estimate $\upDelta_\text{fus} \mu$ by performing Gibbs-Helmholtz integration from the coexistence point ($T_1=\SI{274.6}{\kelvin}$\cite{Molinero2009}). To do this, we can evaluate
\begin{equation}
\frac{G_\text{m}(T_2,\,p)}{T_2} = \text{constant} - \int_{T_1}^{T_2} \frac{\left\langle U\right\rangle_\text{m}+p\left\langle V\right\rangle_\text{m}}{T^2}\,\mathrm{d}T,
\end{equation}
where $\left\langle U\right\rangle_\text{m}$ is the mean molar potential energy and $\left\langle V\right\rangle_\text{m}$ is the mean molar volume measured in a simulation, averaged over a reasonably large number of Monte Carlo cycles. We perform such simulations over a range of temperatures for both the liquid state and the hexagonal ice state and then find best-fit equations for the mean potential energy and the mean volume, and integrate appropriately. Using this approach in a series of short simulations, we find that the chemical potential difference between the two phases is $\upDelta_\text{fus}\mu(\SI{220}{\kelvin})/k_\text{B}\approx\SI{118}{\kelvin}$ per particle, in surprisingly good agreement, considering the degree of supercooling, with the value obtained from the simple approximation that $\upDelta_{\text{fus}} \mu(T) \approx \upDelta_{\text{fus}} H_\mathrm{m} ((T_\text{fus}-T)/T_\text{fus})$,\cite{Oxtoby1992} where we use $\upDelta_\text{fus} H = \SI{1.26}{\kilo\calorie\per\mole}$ and $T_\text{fus}=\SI{274.6}{\kelvin}$,\cite{Molinero2009} giving $\upDelta_\text{fus}\mu(\SI{220}{\kelvin})/k_\text{B}\approx\SI{126}{\kelvin}$ per particle. We are not aware of any calculated values of the interfacial free energy term for the MW potential and so we use that determined for the TIP4P potential, \SI{24}{\milli\joule\per\metre\squared}.\cite{Handel2008} The classical nucleation theory free energy profile is shown for comparison in Fig.~\ref{fig-MW-energybarrier}; it is qualitatively similar to the computed profiles, although the barrier is somewhat higher. One aspect to bear in mind is that the CNT profile passes through the origin, whereas for our order parameter, the size of the largest crystalline cluster has an average value of two in the supercooled water. Interestingly, a line of best fit corresponding to $\upDelta G(N) = -\upDelta_\text{fus}\mu N + t_2 N^{2/3} + t_3 N^{1/3} + t_4$, where $t_i$ are parameters determined by regression ($\upDelta G(N)/k_\text{B}T = -0.54 N+3.45 N^{2/3}+0.84 N^{1/3}-6.13$), suggests the interfacial free energy attributable to the lowest free energy pathway is $\gamma\approx \SI{22.9}{\milli\joule\per\metre\squared}$ (also shown in Fig.~\ref{fig-MW-energybarrier}).\cite{Note7} A similar analysis for the system grown directly from the supercooled liquid, where we fit all four parameters, including $\upDelta_\text{fus} \mu$, suggests that $\upDelta_\text{fus}\mu(\SI{220}{\kelvin})/k_\text{B}\approx\SI{120}{\kelvin}$ and $\gamma\approx \SI{26.2}{\milli\joule\per\metre\squared}$; in other words, the chemical potential does not change significantly, but the interfacial energy becomes somewhat less favourable for `mixed' crystal growth. Given the physically reasonable values obtained from these fits, we can conclude that classical nucleation theory appears to apply to a reasonable degree to the nucleation of MW ice; an accurate numerical determination of the interfacial free energy for the MW model would help in assessing this observation.
It is also interesting to note that the enthalpy associated with the growth of the crystal nucleus is a monotonically downhill function of the nucleus size; this result implies that the barrier to nucleation is primarily entropic in nature.

Li and co-workers estimated the interfacial free energy from a fit of the nucleation rate to the CNT result as a function of the temperature, assuming that $\gamma$ is temperature-independent and that the critical nucleus is spherical. They reported a value of $\gamma=\SI{31.01}{\milli\joule\per\metre\squared}$,\cite{Li2011} which is higher than what we have determined.  In order to compare our results with theirs, we have calculated the order parameter used by Li and co-workers for a large number ($\sim$\num{60000}) of independent configurations taken along the nucleation pathway for the simulations started from the supercooled liquid; we weighted each value with the associated free energy from Fig.~\ref{fig-MW-energybarrier}. This produced an approximate free energy profile corresponding to nucleation measured with their order parameter. Fitting a CNT-like expression, $\upDelta G(N)/k_\text{B}T = t_1 N + t_2 N^{2/3} + t_3 N^{1/3} + t_4$, to this free energy profile allows us to calculate the interfacial free energy associated with the process. If we fix $t_1$ to the bulk chemical potential change ($t_1=-\upDelta_\text{fus}\mu(\SI{220}{\kelvin})/k_\text{B}T$), we find that $\gamma\approx\SI{34.9}{\milli\joule\per\metre\squared}$, in reasonable agreement with the result of Li and co-workers. However, this fit gives $t_3=-8.9$ and $t_4=-0.04$; this is a much larger value of $t_3$ than we obtained with our order parameter, and because $t_3$ does not appear in the CNT expression, this suggests that this may not be the best fit to the data. By contrast, if we fit all four $t_i$ parameters and interpret them using classical nucleation theory, we find that $\upDelta_\text{fus} \mu(\SI{220}{\kelvin})/k_\text{B} \approx \SI{81.52}{\kelvin}$ and $\gamma\approx\SI{20.8}{\milli\joule\per\metre\squared}$; here, $t_3=0.28$, suggesting that the fit to classical nucleation theory is much better.

We suggest that this fitting process may provide a measure of how reasonable the order parameter used to track the nucleation process is; an unconstrained fit should reproduce $\upDelta_\text{fus}\mu$ reasonably when fitting data to CNT. This is the case when using our order parameter, but not when using the order parameter of Li and co-workers; their order parameter is too generous and the result is a chemical potential difference that does not appear to favour ice as much as it should, and a consequently lower interfacial free energy, since much of the ice cluster is liquid-like. If the assumption is that $\upDelta_\text{fus}\mu$ can be approximated by its bulk value in this case, then the interfacial free energy will necessarily be overestimated in compensation. We suggest, therefore, that counting `surface' particles as being part of the ice cluster may result in an order parameter that is not well suited to being utilised in a fit to classical nucleation theory.

Since the MW potential is a reparameterisation of the Stillinger-Weber potential for silicon, it is worthwhile also briefly to compare the free energy profile to that calculated for the nucleation of Stillinger-Weber silicon, which Beaucage and Mousseau simulated in the canonical ensemble at \SI{25}{\percent} supercooling and for which they determined the critical nucleus size to be approximately 175.\cite{Beaucage2005} Although the conditions and the order parameter used to track the nucleation process differ from our simulations, we note that the increased weight given to tetrahedrality in the MW potential appears to result in a smaller critical nucleus size. This is not unexpected in the light of classical nucleation theory, as the greater similarity of the liquid to the solid presumably decreases the interfacial free energy.

\begin{figure}
\centering
\includegraphics{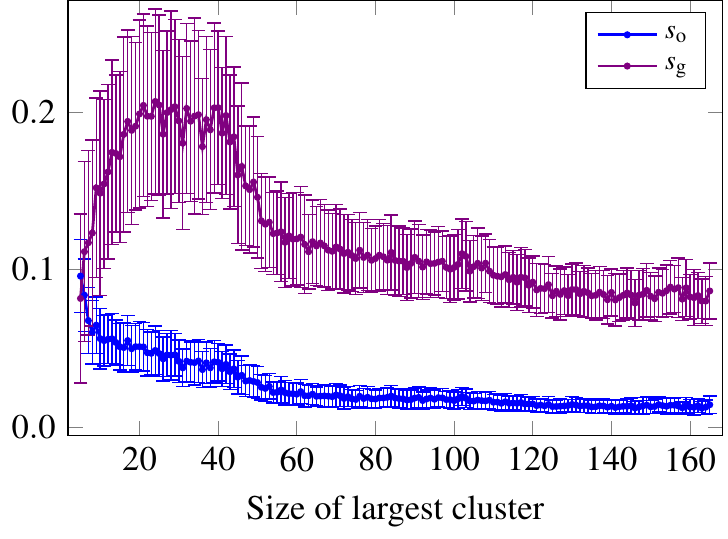}
\caption{A plot of two sphericity parameters against the size of the largest crystalline cluster calculated for snapshots of the system along the local order parameter used to drive nucleation in our system.  Error bars show the standard deviation for the population of configurations at each cluster size. The diagram shows results for nucleation from a supercooled liquid; the other two systems behave analogously.}\label{fig-MW-sphericity}
\end{figure}

From a visual inspection of configurations, the crystalline clusters seemed to be fairly spherical. To quantify this observation, we have computed two measures of the sphericity of the clusters. The first is based upon the moment of inertia tensor, which has elements
\begin{equation}
 I_{ij} = \sum_{k=1}^N m_k (r_k^2 \delta_{ij} - r_{ki} r_{kj}),
\end{equation}
where $r_{ki}$ is the $i$-th component of the vector between the centre of mass and particle $k$, $r_k$ is the overall magnitude of this vector and $\delta_{ij}$ is the Kronecker delta; $i$ and $j$ refer to cartesian co-ordinates in an arbitrary laboratory frame of reference. The eigenvalues of this tensor give the principal components of the moment of inertia; we use these to quantify the orientational sphericity using the asphericity parameter\cite{Rudnick1986}
\begin{equation}
s_\text{o} = \frac{(I_{xx}-I_{yy})^2 + (I_{xx}-I_{zz})^2 + (I_{yy}-I_{zz})^2}{2 (I_{xx} + I_{yy}+I_{zz})^2},
\end{equation}
which ranges between zero for a perfectly spherical cluster and unity for extremely elongated ones.
Any cluster with $s_\text{o} \lessapprox 0.1$ appears very much spherical by visual inspection. As the clusters grow in our simulations, their sphericity increases, as can be seen in Fig.~\ref{fig-MW-sphericity}, although this is to some extent a result of the discrete nature of clusters being felt more at smaller sizes.
We also characterise sphericity in terms of the radius of gyration, namely using\cite{Moore2009}
\begin{equation}
s_\text{g} = \frac{R_\text{g}}{1.5 c^{1/3}}-1, \text{ where } R_\text{g}^2 = \frac{1}{c^2} \sum_{i}^{c} \sum_{j>i}^{c} \left| \mathbold{r}_i-\mathbold{r}_j \right|^2,
\end{equation}
$c$ is the size of the cluster and the $1.5c^{1/3}$ factor reflects the approximate radius of gyration of a perfect sphere of ice at the simulation temperature. This parameter tends to zero for perfect spheres. This measure again confirms that the crystalline clusters are reasonably spherical.

We remark here that while these parameters measure spherical symmetry, they do not principally measure compactness: a cluster with small strings growing out of it on all sides will appear to be more spherically symmetric than a compact oval-shaped cluster, even though we might expect the oval cluster to be more favourable (that is, having a lower surface area to bulk volume ratio). Nevertheless, we have not encountered any especially pathological clusters in our visual inspection of configurations.

\subsection{Nucleation pathway}
In order to compare further the nucleation behaviour we observe to that of Quigley and Rodger,\cite{Quigley2008} we have evaluated the global order parameters that they used along the nucleation pathways computed in this work. We used the Steinhardt-style $Q_6$ and Chau-Hardwick-style\cite{Chau1998,Errington2001} tetrahedrality parameters as defined by Quigley and Rodger,\cite{Quigley2008} including their smoothing function, to ensure comparability of results. We thus have
\begin{equation}
 Q_l = \left( \frac{4\uppi}{2l+1} \sum_{m=-l}^{+l} \left| \frac{1}{4N}  \sum_{i=1}^{N} \sum_{j\ne i}^{N} f(r_{ij}) S_{lm}\left(\theta_{ij},\,\varphi_{ij}\right)  \right|^2  \right)^{1/2},
\end{equation}
where $N$ is the number of particles, $f(r_{ij})$ is the smoothing function, and we again use real spherical harmonics (the results with complex spherical harmonics are, of course, identical),
and\cite{Note8}\nocite{Quigley2009}
\begin{equation}
 \zeta =  \frac{1}{4N} \sum_{i=1}^N \sum_{\substack{j=1\\j\ne i}}^N \sum_{\substack{k>j\\k\ne i}}^{N} f(r_{ij})f(r_{ik}) \left( \hat{\mathbold{r}}_{ij} \cdot \hat{\mathbold{r}}_{ik} + 1/3) \right)^2,\label{eqn-ChauHardwick}
\end{equation}
where $\hat{\mathbold{r}}_{ij}$ is the unit vector from particle $i$ to particle $j$. The smoothing function is defined as
\begin{equation}
f(r) = \begin{cases}
       1 & \quad \text{if } r \le \SI{3.1}{\angstrom}, \\
       \left(\cos\frac{(r/\si{\angstrom}-3.1)\uppi}{0.4}+1\right)/2 & \quad \text{if } \SI{3.1}{\angstrom} < r \le \SI{3.5}{\angstrom}, \\
       0 & \quad \text{otherwise.}
    \end{cases}
\end{equation}
The bulk $\zeta$ parameter for supercooled MW liquid water at \SI{220}{\kelvin} tends to $\sim$0.21. For equilibrated cubic and hexagonal ice at \SI{220}{\kelvin}, $\zeta$ approaches 0.026, while $Q_6$ approaches 0.44 for hexagonal and 0.51 for cubic ice.

\begin{figure}
\centering
\includegraphics{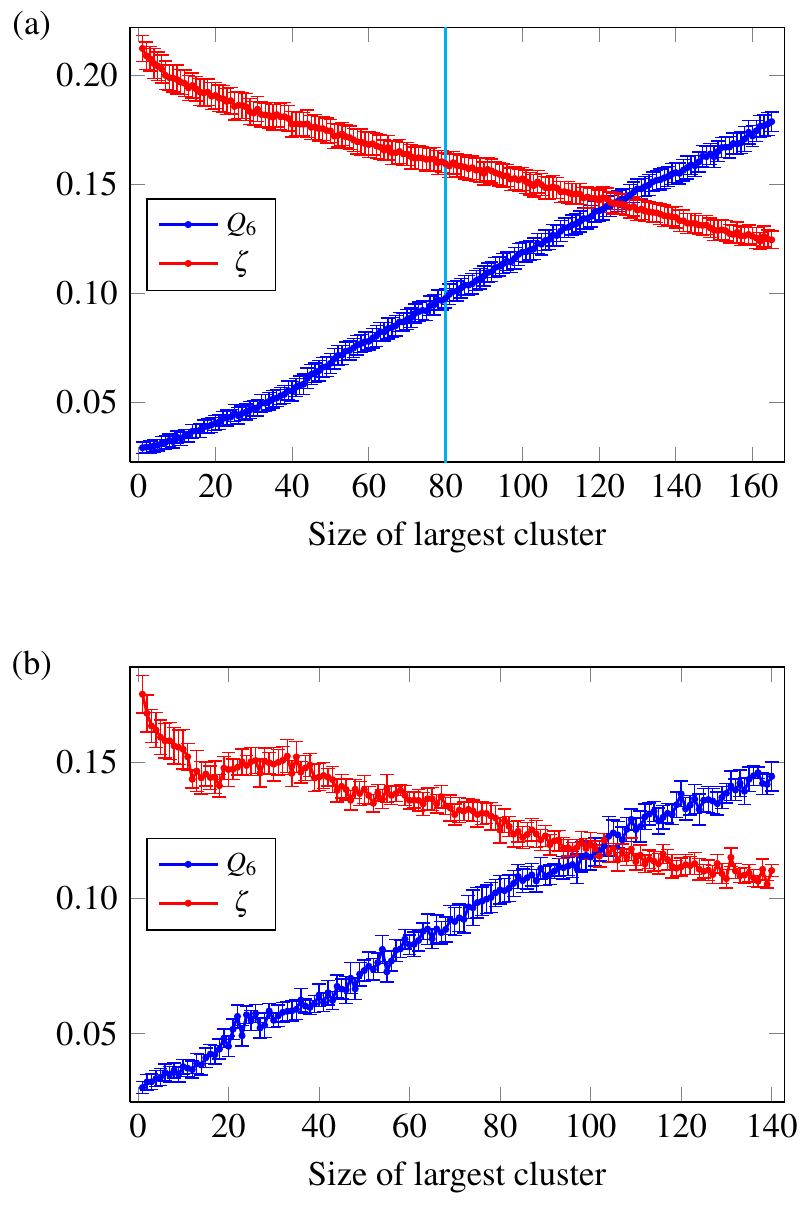}
\caption{(a) The global order parameters $Q_6$ and $\zeta$ calculated along the local order parameter used to drive nucleation in our system. Error bars show the standard deviation for the population of configurations at each cluster size. The cyan solid line indicates the approximate location of the nucleation barrier as calculated by local order parameters. The results depicted here refer to the 576 particles around the centre of mass of the ice nucleus for comparison with Ref.~\onlinecite{Quigley2008}, and not to the full 1400-particle system. (b) An analogous diagram for a single brute-force trajectory at \SI{210}{\kelvin}.}\label{fig-MW-ordParamsOfGrowth}
\end{figure}

It is also important to note that as these global order parameters measure the average order \emph{of the entire system}, an ice cluster of a given size will produce different values of $Q_6$ and $\zeta$ depending on the overall size of the system.\cite{Leyssale2005} Therefore, as we wish to compare our results to Fig.~3 of Ref.~\onlinecite{Quigley2008}, where the number of particles in the simulation was 576, we have calculated the global order parameters in such a way that we have only taken into account the nearest 576 particles from the centre of mass of the crystal nucleus as identified by our local order parameter. The mean values of $Q_6$ and $\zeta$ along our pathway for the nucleation from a cubic ice seed are depicted in Fig.~\ref{fig-MW-ordParamsOfGrowth}(a). At the top of our free energy barrier, we find $Q_6\approx 0.10$ and $\zeta\approx 0.16$. By contrast, Quigley and Rodger report that the saddle point of their two-dimensional free energy landscape occurs at $Q_6\approx 0.22$ and $\zeta\approx 0.19$.\cite{Quigley2008} Furthermore, in our simulations, the order parameters $\zeta$ and $Q_6$ vary roughly linearly as the ice cluster grows, as would be expected for the growth of a crystal nucleus into a relatively unperturbed liquid. By contrast, Quigley and Rodger initially observe an increase in $Q_6$ with little change in $\zeta$. The nucleation pathway we observe therefore appears to differ significantly from the one reported by Quigley and Rodger. The sigmoidal form of their $Q_6$-$\zeta$ free energy landscape suggests their pathway involves an initial increase in the $Q_6$ orientational order of the whole system, not just of a crystal nucleus. This `pre-ordering' then appears to make crystallisation more tractable on computational timescales, even though this is not the natural pathway of nucleation.

We have not been able to nucleate ice in a brute-force simulation at \SI{220}{\kelvin}; however, we assume that the nucleation behaviour does not change significantly between this temperature and other temperatures above the liquid transformation temperature of \SI{202}{\kelvin}.\cite{Moore2011}  An analogous diagram to Fig.~\ref{fig-MW-ordParamsOfGrowth}(a) is shown for a representative brute-force simulation of nucleation at \SI{210}{\kelvin} in Fig.~\ref{fig-MW-ordParamsOfGrowth}(b). Even though the critical nucleus size is different, and the data are significantly more noisy, the basic linear behaviour of $Q_6$ and $\zeta$ remains unchanged. We believe that this brute-force behaviour is a good indication that applying umbrella sampling does not change the natural nucleation pathway.

\subsection{Nucleation rate}
Configurations from the top of the barrier obtained in Monte Carlo umbrella sampling simulations were taken as starting points in isobaric-isothermal MD simulations (with a Nos\'{e}-Hoover barostat and thermostat with relaxation times \SI{2}{\pico\second} and \SI{1}{\pico\second}, respectively, and a time step of \SI{5}{\femto\second}). Velocity components were assigned from a normal distribution with mean zero and scaled to correspond to the simulation temperature of \SI{220}{\kelvin}.

For the system nucleating from a seed hexagonal crystal, we ran 40 simulations from distinct starting configurations at the top of the barrier. Of these, 17 exhibited nucleus growth and the remainder nucleus shrinkage. This roughly $1\,:\,1$ ratio of growing and shrinking suggests that we are approximately at the top of the barrier. From these simulations, we find the long-time attachment rate (averaged over all configurations, not just the ones remaining close to the critical nucleus size\cite{Filion2010}) to be $f_{n_\text{crit}} = \SI{1.175e13}{\per\second}$. The Zeldovich factor is $Z^2=0.000344$ and the number density of the liquid is $\rho_\text{liq} = \SI{3.3e28}{\per\cubic\metre}$. The probability of reaching the top of the free energy barrier is $P(n_\text{crit}) = \exp\left[-\upDelta G(n_\text{crit})/k_\text{B}T\right] = \num{9.0e-9}$. The rate of MW ice nucleation at \SI{220}{\kelvin} is therefore estimated to be $R = \SI{6.5e31}{\per\cubic\metre\per\second}$.

We performed analogous simulations starting from the critical cluster in the system which was grown directly from the supercooled liquid and has a critical nucleus size of 114. Of the 40 MD trajectories started from this point, 20 exhibited shrinking and 20 exhibited growth of the largest crystal nucleus. An equivalent analysis to that above gives a rate of nucleation of $R =  \SI{2.6e29}{\per\cubic\metre\per\second}$.

Using forward flux sampling\cite{Allen2005,Allen2006,Allen2006b} with molecular dynamics,\cite{Wang2009,Li2009b} Li and co-workers have determined\cite{Li2011} nucleation rates for the homogeneous nucleation of ice with the MW model ranging between $\SI{2.148e25}{\per\cubic\metre\per\second}$ at \SI{220}{\kelvin} and $\SI{1.672e-7}{\per\cubic\metre\per\second}$ at \SI{240}{\kelvin}, 5--10 orders of magnitude below the experimental values at the temperatures where these have been reported. Our results are somewhat higher than those reported by Li and co-workers. To an extent, we can rationalise this difference by noting that their use of forward flux sampling ensures a purely kinetically driven process; by contrast, our systems are locally equilibrated. This is not usually a problem with algorithms such as forward flux sampling because local equilibration is generally considerably faster than progression to a higher value of the order parameter serving as a reaction co-ordinate; however, in systems such as water, local equilibration can be very slow. This serves to rationalise the difference in the calculated rate between the system we have grown directly from the supercooled liquid and the one simulated by Li and co-workers.

A second issue is that there is no straightforward way for a simulation method to choose the pathway of fastest growth if this difference arises only for somewhat larger crystal nuclei; indeed, the free energy profiles we have calculated look the same for small clusters. In forward flux sampling, the order parameter space is staged; however, if all the pathways in the initial stages are equally fast, then the pathways chosen in forward flux sampling will be the ones that are the dominant ones in numbers, not the ones that may, ultimately, prove to be faster. Once an ultimately slower pathway is chosen, the simulation is extremely unlikely to be able to backtrack and choose a faster pathway.  We suggest that this is why Li and co-workers did not see the nucleation of a pure phase of ice, even though our results suggest that the rate of nucleation of pure hexagonal ice is two orders of magnitude higher than that of mixed ice. A greedy simulation algorithm such as forward flux sampling, despite sampling natural dynamics, is even more likely to proceed down the `wrong' pathway than simulation methods which allow for local equilibration, such as umbrella sampling. Nevertheless, umbrella sampling is by no means immune to this problem: indeed, the hexagonal crystal that we have grown was only possible because we effectively imposed that nucleation pathway by seeding the system.

\section{Conclusions}
We have simulated the homogeneous nucleation of ice as a function of the size of the largest crystalline cluster using the MW potential. The free energy barriers we have calculated agree reasonably well with classical nucleation theory, and the interfacial free energies extracted from fits to the computed free energy profiles appear to be sensible. We have observed that the crystalline clusters grown directly from the supercooled liquid tend to be cubic, although they have a higher free energy than those clusters grown from ideal crystalline seeds.

A comparison with the free energy landscapes computed in the previous studies of homogeneous ice nucleation\cite{Radhakrishnan2003b,Quigley2008,Brukhno2008} at a similar undercooling ($\sim$\SI{20}{\percent}), albeit with the TIP4P potential, reveals that the free energy barrier calculated in this work is considerably lower.  We suggest that previous simulations were not at equilibrium in path space; that is, the lowest free energy paths between the liquid and the crystal were not located. The high free energy barriers previously reported may be a result of non-local\cite{Radhakrishnan2003b, Quigley2008} and non-rotationally-invariant\cite{Brukhno2008} order parameters used to track the progress of nucleation. We therefore hypothesise more generally that care should be exercised when using global orientational order parameters, because there may be a tendency for the pathways that are easiest to locate with rare event techniques not actually to be those with the lowest free energy, particularly for systems such as water, where crystal growth is slow.

The simulations we have performed allow us to learn some lessons for future simulations of all-atom models of water. If the MW potential is a reasonable model for the energetics of water, as is strongly suggested by the work of Molinero and co-workers,\cite{Molinero2009,Moore2011} the difficulty in simulating water nucleation in all-atom models such as TIP4P may be more due to the slow kinetics of the process than an insurmountable free energy barrier. We suggest that, to test this hypothesis, a study of the homogeneous nucleation of ice using an all-atom water potential and driven by a local order parameter would be a worthwhile, if difficult, endeavour. We are hopeful that the results for the thermodynamics of ice nucleation presented here will be a useful guide for such a study, which will in turn lead to a fuller understanding of the dynamics of ice nucleation.

\begin{acknowledgments}

We would like to thank the Engineering and Physical Sciences Research Council for financial support, and A.~J.~Williamson and F.~Romano for helpful discussions.
\end{acknowledgments}


\begin{thebibliography}{94}%
\makeatletter
\providecommand \@ifxundefined [1]{%
 \@ifx{#1\undefined}
}%
\providecommand \@ifnum [1]{%
 \ifnum #1\expandafter \@firstoftwo
 \else \expandafter \@secondoftwo
 \fi
}%
\providecommand \@ifx [1]{%
 \ifx #1\expandafter \@firstoftwo
 \else \expandafter \@secondoftwo
 \fi
}%
\providecommand \natexlab [1]{#1}%
\providecommand \enquote  [1]{``#1''}%
\providecommand \bibnamefont  [1]{#1}%
\providecommand \bibfnamefont [1]{#1}%
\providecommand \citenamefont [1]{#1}%
\providecommand \href@noop [0]{\@secondoftwo}%
\providecommand \href [0]{\begingroup \@sanitize@url \@href}%
\providecommand \@href[1]{\@@startlink{#1}\@@href}%
\providecommand \@@href[1]{\endgroup#1\@@endlink}%
\providecommand \@sanitize@url [0]{\catcode `\\12\catcode `\$12\catcode
  `\&12\catcode `\#12\catcode `\^12\catcode `\_12\catcode `\%12\relax}%
\providecommand \@@startlink[1]{}%
\providecommand \@@endlink[0]{}%
\providecommand \url  [0]{\begingroup\@sanitize@url \@url }%
\providecommand \@url [1]{\endgroup\@href {#1}{\urlprefix }}%
\providecommand \urlprefix  [0]{URL }%
\providecommand \Eprint [0]{\href }%
\providecommand \doibase [0]{http://dx.doi.org/}%
\providecommand \selectlanguage [0]{\@gobble}%
\providecommand \bibinfo  [0]{\@secondoftwo}%
\providecommand \bibfield  [0]{\@secondoftwo}%
\providecommand \translation [1]{[#1]}%
\providecommand \BibitemOpen [0]{}%
\providecommand \bibitemStop [0]{}%
\providecommand \bibitemNoStop [0]{.\EOS\space}%
\providecommand \EOS [0]{\spacefactor3000\relax}%
\providecommand \BibitemShut  [1]{\csname bibitem#1\endcsname}%
\let\auto@bib@innerbib\@empty
\bibitem [{\citenamefont {Jeffery}\ and\ \citenamefont
  {Austin}(1997)}]{Jeffery1997}%
  \BibitemOpen
  \bibfield  {author} {\bibinfo {author} {\bibfnamefont {C.~A.}\ \bibnamefont
  {Jeffery}}\ and\ \bibinfo {author} {\bibfnamefont {P.~H.}\ \bibnamefont
  {Austin}},\ }\href {\doibase 10.1029/97JD02243} {\bibfield  {journal}
  {\bibinfo  {journal} {J.\ Geophys.\ Res.--Atmos}\ }\textbf {\bibinfo {volume}
  {102}},\ \bibinfo {pages} {25269} (\bibinfo {year} {1997})}\BibitemShut
  {NoStop}%
\bibitem [{\citenamefont {Pruppacher}\ and\ \citenamefont
  {Klett}(2010)}]{Pruppacher1997}%
  \BibitemOpen
  \bibfield  {author} {\bibinfo {author} {\bibfnamefont {H.~R.}\ \bibnamefont
  {Pruppacher}}\ and\ \bibinfo {author} {\bibfnamefont {J.~D.}\ \bibnamefont
  {Klett}},\ }\href {\doibase 10.1007/978-0-306-48100-0} {\emph {\bibinfo
  {title} {Microphysics of clouds and precipitation}}},\ \bibinfo {edition}
  {2nd}\ ed.\ (\bibinfo  {publisher} {Springer},\ \bibinfo {address}
  {Dordrecht},\ \bibinfo {year} {2010})\BibitemShut {NoStop}%
\bibitem [{\citenamefont {Oxtoby}(1998)}]{Oxtoby1998}%
  \BibitemOpen
  \bibfield  {author} {\bibinfo {author} {\bibfnamefont {D.~W.}\ \bibnamefont
  {Oxtoby}},\ }\href {\doibase 10.1021/ar9702278} {\bibfield  {journal}
  {\bibinfo  {journal} {Acc.\ Chem.\ Res.}\ }\textbf {\bibinfo {volume} {31}},\
  \bibinfo {pages} {91} (\bibinfo {year} {1998})}\BibitemShut {NoStop}%
\bibitem [{\citenamefont {Laaksonen}, \citenamefont {Talanquer},\ and\
  \citenamefont {Oxtoby}(1995)}]{Laaksonen1995}%
  \BibitemOpen
  \bibfield  {author} {\bibinfo {author} {\bibfnamefont {A.}~\bibnamefont
  {Laaksonen}}, \bibinfo {author} {\bibfnamefont {V.}~\bibnamefont
  {Talanquer}}, \ and\ \bibinfo {author} {\bibfnamefont {D.~W.}\ \bibnamefont
  {Oxtoby}},\ }\href {\doibase 10.1146/annurev.pc.46.100195.002421} {\bibfield
  {journal} {\bibinfo  {journal} {Annu.\ Rev.\ Phys.\ Chem.}\ }\textbf
  {\bibinfo {volume} {46}},\ \bibinfo {pages} {489} (\bibinfo {year}
  {1995})}\BibitemShut {NoStop}%
\bibitem [{\citenamefont {Auer}\ and\ \citenamefont
  {Frenkel}(2005)}]{Auer2005}%
  \BibitemOpen
  \bibfield  {author} {\bibinfo {author} {\bibfnamefont {S.}~\bibnamefont
  {Auer}}\ and\ \bibinfo {author} {\bibfnamefont {D.}~\bibnamefont {Frenkel}},\
  }\href {\doibase 10.1007/b99429} {\bibfield  {journal} {\bibinfo  {journal}
  {Adv.\ Polym.\ Sci.}\ }\textbf {\bibinfo {volume} {173}},\ \bibinfo {pages}
  {149} (\bibinfo {year} {2005})}\BibitemShut {NoStop}%
\bibitem [{\citenamefont {Anwar}\ and\ \citenamefont {Zahn}(2011)}]{Anwar2011}%
  \BibitemOpen
  \bibfield  {author} {\bibinfo {author} {\bibfnamefont {J.}~\bibnamefont
  {Anwar}}\ and\ \bibinfo {author} {\bibfnamefont {D.}~\bibnamefont {Zahn}},\
  }\href {\doibase 10.1002/ange.201000463} {\bibfield  {journal} {\bibinfo
  {journal} {Angew.~Chem.-Ger.~Edit.}\ }\textbf {\bibinfo {volume} {123}},\
  \bibinfo {pages} {2042} (\bibinfo {year} {2011})}\BibitemShut {NoStop}%
\bibitem [{\citenamefont {Oxtoby}(1992)}]{Oxtoby1992}%
  \BibitemOpen
  \bibfield  {author} {\bibinfo {author} {\bibfnamefont {D.~W.}\ \bibnamefont
  {Oxtoby}},\ }\href {\doibase 10.1088/0953-8984/4/38/001} {\bibfield
  {journal} {\bibinfo  {journal} {J.\ Phys.:\ Cond.\ Matt.}\ }\textbf {\bibinfo
  {volume} {4}},\ \bibinfo {pages} {7627} (\bibinfo {year} {1992})}\BibitemShut
  {NoStop}%
\bibitem [{\citenamefont {Baker}(1997)}]{Baker1997}%
  \BibitemOpen
  \bibfield  {author} {\bibinfo {author} {\bibfnamefont {M.~B.}\ \bibnamefont
  {Baker}},\ }\href {\doibase 10.1126/science.276.5315.1072} {\bibfield
  {journal} {\bibinfo  {journal} {Science}\ }\textbf {\bibinfo {volume}
  {276}},\ \bibinfo {pages} {1072} (\bibinfo {year} {1997})}\BibitemShut
  {NoStop}%
\bibitem [{\citenamefont {Benz}\ \emph {et~al.}(2005)\citenamefont {Benz},
  \citenamefont {Megahed}, \citenamefont {M{\"{o}}hler}, \citenamefont
  {Saathoff}, \citenamefont {Wagner},\ and\ \citenamefont
  {Schurath}}]{Benz2005}%
  \BibitemOpen
  \bibfield  {author} {\bibinfo {author} {\bibfnamefont {S.}~\bibnamefont
  {Benz}}, \bibinfo {author} {\bibfnamefont {K.}~\bibnamefont {Megahed}},
  \bibinfo {author} {\bibfnamefont {O.}~\bibnamefont {M{\"{o}}hler}}, \bibinfo
  {author} {\bibfnamefont {H.}~\bibnamefont {Saathoff}}, \bibinfo {author}
  {\bibfnamefont {R.}~\bibnamefont {Wagner}}, \ and\ \bibinfo {author}
  {\bibfnamefont {U.}~\bibnamefont {Schurath}},\ }\href {\doibase
  10.1016/j.jphotochem.2005.08.026} {\bibfield  {journal} {\bibinfo  {journal}
  {J.\ Photoch.\ Photobio.\ A}\ }\textbf {\bibinfo {volume} {176}},\ \bibinfo
  {pages} {208} (\bibinfo {year} {2005})}\BibitemShut {NoStop}%
\bibitem [{\citenamefont {Hegg}\ and\ \citenamefont {Baker}(2009)}]{Hegg2009}%
  \BibitemOpen
  \bibfield  {author} {\bibinfo {author} {\bibfnamefont {D.~A.}\ \bibnamefont
  {Hegg}}\ and\ \bibinfo {author} {\bibfnamefont {M.~B.}\ \bibnamefont
  {Baker}},\ }\href {\doibase 10.1088/0034-4885/72/5/056801} {\bibfield
  {journal} {\bibinfo  {journal} {Rep.\ Prog.\ Phys.}\ }\textbf {\bibinfo
  {volume} {72}},\ \bibinfo {pages} {056801} (\bibinfo {year}
  {2009})}\BibitemShut {NoStop}%
\bibitem [{\citenamefont {Spichtinger}\ and\ \citenamefont
  {Cziczo}(2010)}]{Spichtinger2010}%
  \BibitemOpen
  \bibfield  {author} {\bibinfo {author} {\bibfnamefont {P.}~\bibnamefont
  {Spichtinger}}\ and\ \bibinfo {author} {\bibfnamefont {D.~J.}\ \bibnamefont
  {Cziczo}},\ }\href {\doibase 10.1029/2009JD012168} {\bibfield  {journal}
  {\bibinfo  {journal} {J.~Geophys.~Res.}\ }\textbf {\bibinfo {volume} {115}},\
  \bibinfo {pages} {D14208} (\bibinfo {year} {2010})}\BibitemShut {NoStop}%
\bibitem [{\citenamefont {Toner}, \citenamefont {Cravalho},\ and\ \citenamefont
  {Karel}(1990)}]{Toner1990}%
  \BibitemOpen
  \bibfield  {author} {\bibinfo {author} {\bibfnamefont {M.}~\bibnamefont
  {Toner}}, \bibinfo {author} {\bibfnamefont {E.~G.}\ \bibnamefont {Cravalho}},
  \ and\ \bibinfo {author} {\bibfnamefont {M.}~\bibnamefont {Karel}},\ }\href
  {\doibase 10.1063/1.345670} {\bibfield  {journal} {\bibinfo  {journal}
  {J.~Appl.\ Phys.}\ }\textbf {\bibinfo {volume} {67}},\ \bibinfo {pages}
  {1582} (\bibinfo {year} {1990})}\BibitemShut {NoStop}%
\bibitem [{\citenamefont {Hagen}, \citenamefont {Anderson},\ and\ \citenamefont
  {Kassner}(1981)}]{Hagen1981}%
  \BibitemOpen
  \bibfield  {author} {\bibinfo {author} {\bibfnamefont {D.~E.}\ \bibnamefont
  {Hagen}}, \bibinfo {author} {\bibfnamefont {R.~J.}\ \bibnamefont {Anderson}},
  \ and\ \bibinfo {author} {\bibfnamefont {J.~L.}\ \bibnamefont {Kassner}},\
  }\href {\doibase 10.1175/1520-0469(1981)038<1236:hcnrmf>2.0.co;2} {\bibfield
  {journal} {\bibinfo  {journal} {J.\ Atmos.\ Sci.}\ }\textbf {\bibinfo
  {volume} {38}},\ \bibinfo {pages} {1236} (\bibinfo {year}
  {1981})}\BibitemShut {NoStop}%
\bibitem [{\citenamefont {Pruppacher}(1995)}]{Pruppacher1995}%
  \BibitemOpen
  \bibfield  {author} {\bibinfo {author} {\bibfnamefont {H.~R.}\ \bibnamefont
  {Pruppacher}},\ }\href {\doibase
  10.1175/1520-0469(1995)052<1924:anlahi>2.0.co;2} {\bibfield  {journal}
  {\bibinfo  {journal} {J.\ Atmos.\ Sci.}\ }\textbf {\bibinfo {volume} {52}},\
  \bibinfo {pages} {1924} (\bibinfo {year} {1995})}\BibitemShut {NoStop}%
\bibitem [{\citenamefont {Koop}\ \emph {et~al.}(2000)\citenamefont {Koop},
  \citenamefont {Luo}, \citenamefont {Tsias},\ and\ \citenamefont
  {Peter}}]{Koop2000}%
  \BibitemOpen
  \bibfield  {author} {\bibinfo {author} {\bibfnamefont {T.}~\bibnamefont
  {Koop}}, \bibinfo {author} {\bibfnamefont {B.}~\bibnamefont {Luo}}, \bibinfo
  {author} {\bibfnamefont {A.}~\bibnamefont {Tsias}}, \ and\ \bibinfo {author}
  {\bibfnamefont {T.}~\bibnamefont {Peter}},\ }\href {\doibase
  10.1038/35020537} {\bibfield  {journal} {\bibinfo  {journal} {Nature}\
  }\textbf {\bibinfo {volume} {406}},\ \bibinfo {pages} {611} (\bibinfo {year}
  {2000})}\BibitemShut {NoStop}%
\bibitem [{\citenamefont {Debenedetti}(2003)}]{Debenedetti2003}%
  \BibitemOpen
  \bibfield  {author} {\bibinfo {author} {\bibfnamefont {P.~G.}\ \bibnamefont
  {Debenedetti}},\ }\href {\doibase 10.1088/0953-8984/15/45/R01} {\bibfield
  {journal} {\bibinfo  {journal} {J.\ Phys.:\ Cond.\ Matt.}\ }\textbf {\bibinfo
  {volume} {15}},\ \bibinfo {pages} {R1669} (\bibinfo {year}
  {2003})}\BibitemShut {NoStop}%
\bibitem [{\citenamefont {Kabath}\ \emph {et~al.}(2006)\citenamefont {Kabath},
  \citenamefont {St{\"{o}}ckel}, \citenamefont {Lindinger},\ and\ \citenamefont
  {Baumg{\"{a}}rtel}}]{Kabath2006}%
  \BibitemOpen
  \bibfield  {author} {\bibinfo {author} {\bibfnamefont {P.}~\bibnamefont
  {Kabath}}, \bibinfo {author} {\bibfnamefont {P.}~\bibnamefont
  {St{\"{o}}ckel}}, \bibinfo {author} {\bibfnamefont {A.}~\bibnamefont
  {Lindinger}}, \ and\ \bibinfo {author} {\bibfnamefont {H.}~\bibnamefont
  {Baumg{\"{a}}rtel}},\ }\href {\doibase 10.1016/j.molliq.2005.11.025}
  {\bibfield  {journal} {\bibinfo  {journal} {J.\ Mol.\ Liq.}\ }\textbf
  {\bibinfo {volume} {125}},\ \bibinfo {pages} {204} (\bibinfo {year}
  {2006})}\BibitemShut {NoStop}%
\bibitem [{\citenamefont {Kr{\"{a}}mer}\ \emph {et~al.}(1999)\citenamefont
  {Kr{\"{a}}mer}, \citenamefont {H{\"{u}}bner}, \citenamefont {Vortisch},
  \citenamefont {W{\"{o}}ste}, \citenamefont {Leisner}, \citenamefont
  {Schwell}, \citenamefont {R{\"{u}}hl},\ and\ \citenamefont
  {Baumg{\"{a}}rtel}}]{Kraemer1999}%
  \BibitemOpen
  \bibfield  {author} {\bibinfo {author} {\bibfnamefont {B.}~\bibnamefont
  {Kr{\"{a}}mer}}, \bibinfo {author} {\bibfnamefont {O.}~\bibnamefont
  {H{\"{u}}bner}}, \bibinfo {author} {\bibfnamefont {H.}~\bibnamefont
  {Vortisch}}, \bibinfo {author} {\bibfnamefont {L.}~\bibnamefont
  {W{\"{o}}ste}}, \bibinfo {author} {\bibfnamefont {T.}~\bibnamefont
  {Leisner}}, \bibinfo {author} {\bibfnamefont {M.}~\bibnamefont {Schwell}},
  \bibinfo {author} {\bibfnamefont {E.}~\bibnamefont {R{\"{u}}hl}}, \ and\
  \bibinfo {author} {\bibfnamefont {H.}~\bibnamefont {Baumg{\"{a}}rtel}},\
  }\href {\doibase 10.1063/1.479946} {\bibfield  {journal} {\bibinfo  {journal}
  {J.~Chem.\ Phys.}\ }\textbf {\bibinfo {volume} {111}},\ \bibinfo {pages}
  {6521} (\bibinfo {year} {1999})}\BibitemShut {NoStop}%
\bibitem [{\citenamefont {Bernal}\ and\ \citenamefont
  {Fowler}(1933)}]{Bernal1933}%
  \BibitemOpen
  \bibfield  {author} {\bibinfo {author} {\bibfnamefont {J.~D.}\ \bibnamefont
  {Bernal}}\ and\ \bibinfo {author} {\bibfnamefont {R.~H.}\ \bibnamefont
  {Fowler}},\ }\href {\doibase 10.1063/1.1749327} {\bibfield  {journal}
  {\bibinfo  {journal} {J.~Chem.\ Phys.}\ }\textbf {\bibinfo {volume} {1}},\
  \bibinfo {pages} {515} (\bibinfo {year} {1933})}\BibitemShut {NoStop}%
\bibitem [{\citenamefont {Pauling}(1935)}]{Pauling1935}%
  \BibitemOpen
  \bibfield  {author} {\bibinfo {author} {\bibfnamefont {L.}~\bibnamefont
  {Pauling}},\ }\href {\doibase 10.1021/ja01315a102} {\bibfield  {journal}
  {\bibinfo  {journal} {J.~Am.\ Chem.\ Soc.}\ }\textbf {\bibinfo {volume}
  {57}},\ \bibinfo {pages} {2680} (\bibinfo {year} {1935})}\BibitemShut
  {NoStop}%
\bibitem [{\citenamefont {Petrenko}\ and\ \citenamefont
  {Whitworth}(1999)}]{Petrenko1999}%
  \BibitemOpen
  \bibfield  {author} {\bibinfo {author} {\bibfnamefont {V.~F.}\ \bibnamefont
  {Petrenko}}\ and\ \bibinfo {author} {\bibfnamefont {R.~W.}\ \bibnamefont
  {Whitworth}},\ }\href@noop {} {\emph {\bibinfo {title} {Physics of ice}}}\
  (\bibinfo  {publisher} {Oxford University Press},\ \bibinfo {address}
  {Oxford},\ \bibinfo {year} {1999})\BibitemShut {NoStop}%
\bibitem [{\citenamefont {Murray}, \citenamefont {Knopf},\ and\ \citenamefont
  {Bertram}(2005)}]{Murray2005}%
  \BibitemOpen
  \bibfield  {author} {\bibinfo {author} {\bibfnamefont {B.~J.}\ \bibnamefont
  {Murray}}, \bibinfo {author} {\bibfnamefont {D.~A.}\ \bibnamefont {Knopf}}, \
  and\ \bibinfo {author} {\bibfnamefont {A.~K.}\ \bibnamefont {Bertram}},\
  }\href {\doibase 10.1038/nature03403} {\bibfield  {journal} {\bibinfo
  {journal} {Nature}\ }\textbf {\bibinfo {volume} {434}},\ \bibinfo {pages}
  {202} (\bibinfo {year} {2005})}\BibitemShut {NoStop}%
\bibitem [{\citenamefont {Peter}\ \emph {et~al.}(2006)\citenamefont {Peter},
  \citenamefont {Marcolli}, \citenamefont {Spichtinger}, \citenamefont {Corti},
  \citenamefont {Baker},\ and\ \citenamefont {Koop}}]{Peter2006}%
  \BibitemOpen
  \bibfield  {author} {\bibinfo {author} {\bibfnamefont {T.}~\bibnamefont
  {Peter}}, \bibinfo {author} {\bibfnamefont {C.}~\bibnamefont {Marcolli}},
  \bibinfo {author} {\bibfnamefont {P.}~\bibnamefont {Spichtinger}}, \bibinfo
  {author} {\bibfnamefont {T.}~\bibnamefont {Corti}}, \bibinfo {author}
  {\bibfnamefont {M.~B.}\ \bibnamefont {Baker}}, \ and\ \bibinfo {author}
  {\bibfnamefont {T.}~\bibnamefont {Koop}},\ }\href {\doibase
  10.1126/science.1135199} {\bibfield  {journal} {\bibinfo  {journal}
  {Science}\ }\textbf {\bibinfo {volume} {314}},\ \bibinfo {pages} {1399}
  (\bibinfo {year} {2006})}\BibitemShut {NoStop}%
\bibitem [{\citenamefont {Shilling}\ \emph {et~al.}(2006)\citenamefont
  {Shilling}, \citenamefont {Tolbert}, \citenamefont {Toon}, \citenamefont
  {Jensen}, \citenamefont {Murray},\ and\ \citenamefont
  {Bertram}}]{Shilling2006}%
  \BibitemOpen
  \bibfield  {author} {\bibinfo {author} {\bibfnamefont {J.~E.}\ \bibnamefont
  {Shilling}}, \bibinfo {author} {\bibfnamefont {M.~A.}\ \bibnamefont
  {Tolbert}}, \bibinfo {author} {\bibfnamefont {O.~B.}\ \bibnamefont {Toon}},
  \bibinfo {author} {\bibfnamefont {E.~J.}\ \bibnamefont {Jensen}}, \bibinfo
  {author} {\bibfnamefont {B.~J.}\ \bibnamefont {Murray}}, \ and\ \bibinfo
  {author} {\bibfnamefont {A.~K.}\ \bibnamefont {Bertram}},\ }\href {\doibase
  10.1029/2006GL026671} {\bibfield  {journal} {\bibinfo  {journal} {Geophys.\
  Res.\ Lett.}\ }\textbf {\bibinfo {volume} {33}},\ \bibinfo {pages} {L17801}
  (\bibinfo {year} {2006})}\BibitemShut {NoStop}%
\bibitem [{\citenamefont {Heymsfield}\ \emph {et~al.}(2011)\citenamefont
  {Heymsfield}, \citenamefont {Thompson}, \citenamefont {Morrison},
  \citenamefont {Bansemer}, \citenamefont {Rasmussen}, \citenamefont {Minnis},
  \citenamefont {Wang},\ and\ \citenamefont {Zhang}}]{Heymsfield2011}%
  \BibitemOpen
  \bibfield  {author} {\bibinfo {author} {\bibfnamefont {A.~J.}\ \bibnamefont
  {Heymsfield}}, \bibinfo {author} {\bibfnamefont {G.}~\bibnamefont
  {Thompson}}, \bibinfo {author} {\bibfnamefont {H.}~\bibnamefont {Morrison}},
  \bibinfo {author} {\bibfnamefont {A.}~\bibnamefont {Bansemer}}, \bibinfo
  {author} {\bibfnamefont {R.~M.}\ \bibnamefont {Rasmussen}}, \bibinfo {author}
  {\bibfnamefont {P.}~\bibnamefont {Minnis}}, \bibinfo {author} {\bibfnamefont
  {Z.}~\bibnamefont {Wang}}, \ and\ \bibinfo {author} {\bibfnamefont
  {D.}~\bibnamefont {Zhang}},\ }\href {\doibase 10.1126/science.1202851}
  {\bibfield  {journal} {\bibinfo  {journal} {Science}\ }\textbf {\bibinfo
  {volume} {333}},\ \bibinfo {pages} {77} (\bibinfo {year} {2011})}\BibitemShut
  {NoStop}%
\bibitem [{\citenamefont {Jorgensen}\ \emph {et~al.}(1983)\citenamefont
  {Jorgensen}, \citenamefont {Chandrasekhar}, \citenamefont {Madura},
  \citenamefont {Impey},\ and\ \citenamefont {Klein}}]{Jorgensen1983}%
  \BibitemOpen
  \bibfield  {author} {\bibinfo {author} {\bibfnamefont {W.~L.}\ \bibnamefont
  {Jorgensen}}, \bibinfo {author} {\bibfnamefont {J.}~\bibnamefont
  {Chandrasekhar}}, \bibinfo {author} {\bibfnamefont {J.~D.}\ \bibnamefont
  {Madura}}, \bibinfo {author} {\bibfnamefont {R.~W.}\ \bibnamefont {Impey}}, \
  and\ \bibinfo {author} {\bibfnamefont {M.~L.}\ \bibnamefont {Klein}},\ }\href
  {\doibase 10.1063/1.445869} {\bibfield  {journal} {\bibinfo  {journal}
  {J.~Chem.\ Phys.}\ }\textbf {\bibinfo {volume} {79}},\ \bibinfo {pages} {926}
  (\bibinfo {year} {1983})}\BibitemShut {NoStop}%
\bibitem [{\citenamefont {Fennell}\ and\ \citenamefont
  {Gezelter}(2005)}]{Fennell2005}%
  \BibitemOpen
  \bibfield  {author} {\bibinfo {author} {\bibfnamefont {C.~J.}\ \bibnamefont
  {Fennell}}\ and\ \bibinfo {author} {\bibfnamefont {J.~D.}\ \bibnamefont
  {Gezelter}},\ }\href {\doibase 10.1021/ct050005s} {\bibfield  {journal}
  {\bibinfo  {journal} {J.\ Chem.\ Theory Comput.}\ }\textbf {\bibinfo {volume}
  {1}},\ \bibinfo {pages} {662} (\bibinfo {year} {2005})}\BibitemShut {NoStop}%
\bibitem [{\citenamefont {Vega}\ \emph {et~al.}(2008)\citenamefont {Vega},
  \citenamefont {Sanz}, \citenamefont {Abascal},\ and\ \citenamefont
  {Noya}}]{Vega2008}%
  \BibitemOpen
  \bibfield  {author} {\bibinfo {author} {\bibfnamefont {C.}~\bibnamefont
  {Vega}}, \bibinfo {author} {\bibfnamefont {E.}~\bibnamefont {Sanz}}, \bibinfo
  {author} {\bibfnamefont {J.~L.~F.}\ \bibnamefont {Abascal}}, \ and\ \bibinfo
  {author} {\bibfnamefont {E.~G.}\ \bibnamefont {Noya}},\ }\href {\doibase
  10.1088/0953-8984/20/15/153101} {\bibfield  {journal} {\bibinfo  {journal}
  {J.\ Phys.:\ Cond.\ Matt.}\ }\textbf {\bibinfo {volume} {20}},\ \bibinfo
  {pages} {153101} (\bibinfo {year} {2008})}\BibitemShut {NoStop}%
\bibitem [{\citenamefont {Bartell}\ and\ \citenamefont
  {Huang}(1994)}]{Bartell1994}%
  \BibitemOpen
  \bibfield  {author} {\bibinfo {author} {\bibfnamefont {L.~S.}\ \bibnamefont
  {Bartell}}\ and\ \bibinfo {author} {\bibfnamefont {J.}~\bibnamefont
  {Huang}},\ }\href {\doibase 10.1021/j100082a011} {\bibfield  {journal}
  {\bibinfo  {journal} {J.~Phys.\ Chem.}\ }\textbf {\bibinfo {volume} {98}},\
  \bibinfo {pages} {7455} (\bibinfo {year} {1994})}\BibitemShut {NoStop}%
\bibitem [{\citenamefont {Murray}\ and\ \citenamefont
  {Bertram}(2006)}]{Murray2006}%
  \BibitemOpen
  \bibfield  {author} {\bibinfo {author} {\bibfnamefont {B.~J.}\ \bibnamefont
  {Murray}}\ and\ \bibinfo {author} {\bibfnamefont {A.~K.}\ \bibnamefont
  {Bertram}},\ }\href {\doibase 10.1039/b513480c} {\bibfield  {journal}
  {\bibinfo  {journal} {Phys.\ Chem.\ Chem.\ Phys.}\ }\textbf {\bibinfo
  {volume} {8}},\ \bibinfo {pages} {186} (\bibinfo {year} {2006})}\BibitemShut
  {NoStop}%
\bibitem [{\citenamefont {Svishchev}\ and\ \citenamefont
  {Kusalik}(1994)}]{Svishchev1994}%
  \BibitemOpen
  \bibfield  {author} {\bibinfo {author} {\bibfnamefont {I.~M.}\ \bibnamefont
  {Svishchev}}\ and\ \bibinfo {author} {\bibfnamefont {P.~G.}\ \bibnamefont
  {Kusalik}},\ }\href {\doibase 10.1103/PhysRevLett.73.975} {\bibfield
  {journal} {\bibinfo  {journal} {Phys.\ Rev.\ Lett.}\ }\textbf {\bibinfo
  {volume} {73}},\ \bibinfo {pages} {975} (\bibinfo {year} {1994})}\BibitemShut
  {NoStop}%
\bibitem [{\citenamefont {Svishchev}\ and\ \citenamefont
  {Kusalik}(1996)}]{Svishchev1996}%
  \BibitemOpen
  \bibfield  {author} {\bibinfo {author} {\bibfnamefont {I.~M.}\ \bibnamefont
  {Svishchev}}\ and\ \bibinfo {author} {\bibfnamefont {P.~G.}\ \bibnamefont
  {Kusalik}},\ }\href {\doibase 10.1021/ja951624l} {\bibfield  {journal}
  {\bibinfo  {journal} {J.~Am.\ Chem.\ Soc.}\ }\textbf {\bibinfo {volume}
  {118}},\ \bibinfo {pages} {649} (\bibinfo {year} {1996})}\BibitemShut
  {NoStop}%
\bibitem [{\citenamefont {Yamada}\ \emph {et~al.}(2002)\citenamefont {Yamada},
  \citenamefont {Mossa}, \citenamefont {Stanley},\ and\ \citenamefont
  {Sciortino}}]{Yamada2002}%
  \BibitemOpen
  \bibfield  {author} {\bibinfo {author} {\bibfnamefont {M.}~\bibnamefont
  {Yamada}}, \bibinfo {author} {\bibfnamefont {S.}~\bibnamefont {Mossa}},
  \bibinfo {author} {\bibfnamefont {H.~E.}\ \bibnamefont {Stanley}}, \ and\
  \bibinfo {author} {\bibfnamefont {F.}~\bibnamefont {Sciortino}},\ }\href
  {\doibase 10.1103/PhysRevLett.88.195701} {\bibfield  {journal} {\bibinfo
  {journal} {Phys.\ Rev.\ Lett.}\ }\textbf {\bibinfo {volume} {88}},\ \bibinfo
  {pages} {195701} (\bibinfo {year} {2002})}\BibitemShut {NoStop}%
\bibitem [{\citenamefont {Matsumoto}, \citenamefont {Saito},\ and\
  \citenamefont {Ohmine}(2002)}]{Matsumoto2002}%
  \BibitemOpen
  \bibfield  {author} {\bibinfo {author} {\bibfnamefont {M.}~\bibnamefont
  {Matsumoto}}, \bibinfo {author} {\bibfnamefont {S.}~\bibnamefont {Saito}}, \
  and\ \bibinfo {author} {\bibfnamefont {I.}~\bibnamefont {Ohmine}},\ }\href
  {\doibase 10.1038/416409a} {\bibfield  {journal} {\bibinfo  {journal}
  {Nature}\ }\textbf {\bibinfo {volume} {416}},\ \bibinfo {pages} {409}
  (\bibinfo {year} {2002})}\BibitemShut {NoStop}%
\bibitem [{\citenamefont {Nada}\ and\ \citenamefont {van~der
  Eerden}(2003)}]{Nada2003}%
  \BibitemOpen
  \bibfield  {author} {\bibinfo {author} {\bibfnamefont {H.}~\bibnamefont
  {Nada}}\ and\ \bibinfo {author} {\bibfnamefont {J.~P. J.~M.}\ \bibnamefont
  {van~der Eerden}},\ }\href {\doibase 10.1063/1.1562610} {\bibfield  {journal}
  {\bibinfo  {journal} {J.~Chem.\ Phys.}\ }\textbf {\bibinfo {volume} {118}},\
  \bibinfo {pages} {7401} (\bibinfo {year} {2003})}\BibitemShut {NoStop}%
\bibitem [{\citenamefont {Radhakrishnan}\ and\ \citenamefont
  {Trout}(2003{\natexlab{a}})}]{Radhakrishnan2003b}%
  \BibitemOpen
  \bibfield  {author} {\bibinfo {author} {\bibfnamefont {R.}~\bibnamefont
  {Radhakrishnan}}\ and\ \bibinfo {author} {\bibfnamefont {B.~L.}\ \bibnamefont
  {Trout}},\ }\href {\doibase 10.1021/ja0211252} {\bibfield  {journal}
  {\bibinfo  {journal} {J.~Am.\ Chem.\ Soc.}\ }\textbf {\bibinfo {volume}
  {125}},\ \bibinfo {pages} {7743} (\bibinfo {year}
  {2003}{\natexlab{a}})}\BibitemShut {NoStop}%
\bibitem [{\citenamefont {Radhakrishnan}\ and\ \citenamefont
  {Trout}(2003{\natexlab{b}})}]{Radhakrishnan2003}%
  \BibitemOpen
  \bibfield  {author} {\bibinfo {author} {\bibfnamefont {R.}~\bibnamefont
  {Radhakrishnan}}\ and\ \bibinfo {author} {\bibfnamefont {B.~L.}\ \bibnamefont
  {Trout}},\ }\href {\doibase 10.1103/PhysRevLett.90.158301} {\bibfield
  {journal} {\bibinfo  {journal} {Phys.\ Rev.\ Lett.}\ }\textbf {\bibinfo
  {volume} {90}},\ \bibinfo {pages} {158301} (\bibinfo {year}
  {2003}{\natexlab{b}})}\BibitemShut {NoStop}%
\bibitem [{\citenamefont {Garc\'{\i}a~Fern\'{a}ndez}, \citenamefont {Abascal},\
  and\ \citenamefont {Vega}(2006)}]{Fernandez2006}%
  \BibitemOpen
  \bibfield  {author} {\bibinfo {author} {\bibfnamefont {R.}~\bibnamefont
  {Garc\'{\i}a~Fern\'{a}ndez}}, \bibinfo {author} {\bibfnamefont {J.~L.~F.}\
  \bibnamefont {Abascal}}, \ and\ \bibinfo {author} {\bibfnamefont
  {C.}~\bibnamefont {Vega}},\ }\href {\doibase 10.1063/1.2183308} {\bibfield
  {journal} {\bibinfo  {journal} {J.~Chem.\ Phys.}\ }\textbf {\bibinfo {volume}
  {124}},\ \bibinfo {pages} {144506} (\bibinfo {year} {2006})}\BibitemShut
  {NoStop}%
\bibitem [{\citenamefont {Vrbka}\ and\ \citenamefont
  {Jungwirth}(2006)}]{Vrbka2006}%
  \BibitemOpen
  \bibfield  {author} {\bibinfo {author} {\bibfnamefont {L.}~\bibnamefont
  {Vrbka}}\ and\ \bibinfo {author} {\bibfnamefont {P.}~\bibnamefont
  {Jungwirth}},\ }\href {\doibase 10.1021/jp064021c} {\bibfield  {journal}
  {\bibinfo  {journal} {J.~Phys.\ Chem.~B}\ }\textbf {\bibinfo {volume}
  {110}},\ \bibinfo {pages} {18126} (\bibinfo {year} {2006})}\BibitemShut
  {NoStop}%
\bibitem [{\citenamefont {Vrbka}\ and\ \citenamefont
  {Jungwirth}(2007)}]{Vrbka2007}%
  \BibitemOpen
  \bibfield  {author} {\bibinfo {author} {\bibfnamefont {L.}~\bibnamefont
  {Vrbka}}\ and\ \bibinfo {author} {\bibfnamefont {P.}~\bibnamefont
  {Jungwirth}},\ }\href {\doibase 10.1016/j.molliq.2006.12.011} {\bibfield
  {journal} {\bibinfo  {journal} {J.\ Mol.\ Liq.}\ }\textbf {\bibinfo {volume}
  {134}},\ \bibinfo {pages} {64} (\bibinfo {year} {2007})}\BibitemShut
  {NoStop}%
\bibitem [{\citenamefont {Carignano}(2007)}]{Carignano2007}%
  \BibitemOpen
  \bibfield  {author} {\bibinfo {author} {\bibfnamefont {M.~A.}\ \bibnamefont
  {Carignano}},\ }\href {\doibase 10.1021/jp067388q} {\bibfield  {journal}
  {\bibinfo  {journal} {J.~Phys.\ Chem.~C}\ }\textbf {\bibinfo {volume}
  {111}},\ \bibinfo {pages} {501} (\bibinfo {year} {2007})}\BibitemShut
  {NoStop}%
\bibitem [{\citenamefont {Quigley}\ and\ \citenamefont
  {Rodger}(2008)}]{Quigley2008}%
  \BibitemOpen
  \bibfield  {author} {\bibinfo {author} {\bibfnamefont {D.}~\bibnamefont
  {Quigley}}\ and\ \bibinfo {author} {\bibfnamefont {P.~M.}\ \bibnamefont
  {Rodger}},\ }\href {\doibase 10.1063/1.2888999} {\bibfield  {journal}
  {\bibinfo  {journal} {J.~Chem.\ Phys.}\ }\textbf {\bibinfo {volume} {128}},\
  \bibinfo {pages} {154518} (\bibinfo {year} {2008})}\BibitemShut {NoStop}%
\bibitem [{\citenamefont {Brukhno}\ \emph {et~al.}(2008)\citenamefont
  {Brukhno}, \citenamefont {Anwar}, \citenamefont {Davidchack},\ and\
  \citenamefont {Handel}}]{Brukhno2008}%
  \BibitemOpen
  \bibfield  {author} {\bibinfo {author} {\bibfnamefont {A.~V.}\ \bibnamefont
  {Brukhno}}, \bibinfo {author} {\bibfnamefont {J.}~\bibnamefont {Anwar}},
  \bibinfo {author} {\bibfnamefont {R.}~\bibnamefont {Davidchack}}, \ and\
  \bibinfo {author} {\bibfnamefont {R.}~\bibnamefont {Handel}},\ }\href
  {\doibase 10.1088/0953-8984/20/49/494243} {\bibfield  {journal} {\bibinfo
  {journal} {J.\ Phys.:\ Cond.\ Matt.}\ }\textbf {\bibinfo {volume} {20}},\
  \bibinfo {pages} {494243} (\bibinfo {year} {2008})}\BibitemShut {NoStop}%
\bibitem [{\citenamefont {Pluha\v{r}ov\'{a}}, \citenamefont {Vrbka},\ and\
  \citenamefont {Jungwirth}(2010)}]{Pluharova2010}%
  \BibitemOpen
  \bibfield  {author} {\bibinfo {author} {\bibfnamefont {E.}~\bibnamefont
  {Pluha\v{r}ov\'{a}}}, \bibinfo {author} {\bibfnamefont {L.}~\bibnamefont
  {Vrbka}}, \ and\ \bibinfo {author} {\bibfnamefont {P.}~\bibnamefont
  {Jungwirth}},\ }\href {\doibase 10.1021/jp9090238} {\bibfield  {journal}
  {\bibinfo  {journal} {J.~Phys.~Chem.~B}\ }\textbf {\bibinfo {volume} {114}},\
  \bibinfo {pages} {7831} (\bibinfo {year} {2010})}\BibitemShut {NoStop}%
\bibitem [{\citenamefont {Pereyra}, \citenamefont {Szleifer},\ and\
  \citenamefont {Carignano}(2011)}]{Pereyra2011}%
  \BibitemOpen
  \bibfield  {author} {\bibinfo {author} {\bibfnamefont {R.~G.}\ \bibnamefont
  {Pereyra}}, \bibinfo {author} {\bibfnamefont {I.}~\bibnamefont {Szleifer}}, \
  and\ \bibinfo {author} {\bibfnamefont {M.~A.}\ \bibnamefont {Carignano}},\
  }\href {\doibase 10.1063/1.3613672} {\bibfield  {journal} {\bibinfo
  {journal} {J.~Chem.\ Phys.}\ }\textbf {\bibinfo {volume} {135}},\ \bibinfo
  {pages} {034508} (\bibinfo {year} {2011})}\BibitemShut {NoStop}%
\bibitem [{\citenamefont {Weiss}\ \emph {et~al.}(2011)\citenamefont {Weiss},
  \citenamefont {Rullich}, \citenamefont {K\"{o}hler},\ and\ \citenamefont
  {Frauenheim}}]{Weiss2011}%
  \BibitemOpen
  \bibfield  {author} {\bibinfo {author} {\bibfnamefont {V.~C.}\ \bibnamefont
  {Weiss}}, \bibinfo {author} {\bibfnamefont {M.}~\bibnamefont {Rullich}},
  \bibinfo {author} {\bibfnamefont {C.}~\bibnamefont {K\"{o}hler}}, \ and\
  \bibinfo {author} {\bibfnamefont {T.}~\bibnamefont {Frauenheim}},\ }\href
  {\doibase 10.1063/1.3609768} {\bibfield  {journal} {\bibinfo  {journal}
  {J.~Chem.\ Phys.}\ }\textbf {\bibinfo {volume} {135}},\ \bibinfo {pages}
  {034701} (\bibinfo {year} {2011})}\BibitemShut {NoStop}%
\bibitem [{\citenamefont {Rozmanov}\ and\ \citenamefont
  {Kusalik}(2011)}]{Rozmanov2011}%
  \BibitemOpen
  \bibfield  {author} {\bibinfo {author} {\bibfnamefont {D.}~\bibnamefont
  {Rozmanov}}\ and\ \bibinfo {author} {\bibfnamefont {P.~G.}\ \bibnamefont
  {Kusalik}},\ }\href {\doibase 10.1039/C1CP21210A} {\bibfield  {journal}
  {\bibinfo  {journal} {Phys.\ Chem.\ Chem.\ Phys.}\ }\textbf {\bibinfo
  {volume} {13}},\ \bibinfo {pages} {15501} (\bibinfo {year}
  {2011})}\BibitemShut {NoStop}%
\bibitem [{\citenamefont {Steinhardt}, \citenamefont {Nelson},\ and\
  \citenamefont {Ronchetti}(1983)}]{Steinhardt1983}%
  \BibitemOpen
  \bibfield  {author} {\bibinfo {author} {\bibfnamefont {P.~J.}\ \bibnamefont
  {Steinhardt}}, \bibinfo {author} {\bibfnamefont {D.~R.}\ \bibnamefont
  {Nelson}}, \ and\ \bibinfo {author} {\bibfnamefont {M.}~\bibnamefont
  {Ronchetti}},\ }\href {\doibase 10.1103/PhysRevB.28.784} {\bibfield
  {journal} {\bibinfo  {journal} {Phys.\ Rev.\ B}\ }\textbf {\bibinfo {volume}
  {28}},\ \bibinfo {pages} {784} (\bibinfo {year} {1983})}\BibitemShut
  {NoStop}%
\bibitem [{\citenamefont {ten Wolde}, \citenamefont {Ruiz-Montero},\ and\
  \citenamefont {Frenkel}(1996)}]{TenWolde1996}%
  \BibitemOpen
  \bibfield  {author} {\bibinfo {author} {\bibfnamefont {P.-R.}\ \bibnamefont
  {ten Wolde}}, \bibinfo {author} {\bibfnamefont {M.~J.}\ \bibnamefont
  {Ruiz-Montero}}, \ and\ \bibinfo {author} {\bibfnamefont {D.}~\bibnamefont
  {Frenkel}},\ }\href {\doibase 10.1039/FD9960400093} {\bibfield  {journal}
  {\bibinfo  {journal} {Faraday Discuss.}\ }\textbf {\bibinfo {volume} {104}},\
  \bibinfo {pages} {93} (\bibinfo {year} {1996})}\BibitemShut {NoStop}%
\bibitem [{Note1()}]{Note1}%
  \BibitemOpen
  \bibinfo {note} {The actual pressures used are \SI {1}{atm}\cite
  {Quigley2008} and \SI {1}{\barunit },\cite {Radhakrishnan2003b, Brukhno2008}
  but they are for all intents and purposes equivalent within the simulation
  error.}\BibitemShut {Stop}%
\bibitem [{Note2()}]{Note2}%
  \BibitemOpen
  \bibinfo {note} {We note here that the system size in Quigley and Rodger's
  simulation is in the range where a cylindrical cluster that spans the box has
  the lowest surface to volume ratio for ice clusters at the top of the
  barrier.\cite {Quigley2008} However, this can only act to lower the free
  energy barrier compared to a spherical critical cluster.}\BibitemShut {Stop}%
\bibitem [{Note3()}]{Note3}%
  \BibitemOpen
  \bibinfo {note} {The enthalpy of fusion for the TIP4P model is $\upDelta
  _\protect \text {fus} H = \SI {1.05}{\kilo \calorie \per \mole }$;\cite
  {Vega2005} we assume the density of ice is only slightly increased at \SI
  {180}{\kelvin } at $\rho (\protect \text {I}_\protect \text {h})=\SI
  {0.945}{\gram \per \milli \liter }$ from the melting point density of $\rho
  (\protect \text {I}_\protect \text {h})=\SI {0.940}{\gram \per \milli \liter
  }$;\cite {Vega2005} the melting temperature is $T_\protect \text {fus}=\SI
  {232}{\kelvin }$;\cite {Vega2005} and we use the measured interfacial free
  energy of $\gamma =\SI {24}{\milli \joule \per \metre \squared }$ (calculated
  for TIP4P water with a sharp cutoff, rather than with Ewald summation).\cite
  {Handel2008}}\BibitemShut {NoStop}%
\bibitem [{\citenamefont {Vega}, \citenamefont {Sanz},\ and\ \citenamefont
  {Abascal}(2005)}]{Vega2005}%
  \BibitemOpen
  \bibfield  {author} {\bibinfo {author} {\bibfnamefont {C.}~\bibnamefont
  {Vega}}, \bibinfo {author} {\bibfnamefont {E.}~\bibnamefont {Sanz}}, \ and\
  \bibinfo {author} {\bibfnamefont {J.~L.~F.}\ \bibnamefont {Abascal}},\ }\href
  {\doibase 10.1063/1.1862245} {\bibfield  {journal} {\bibinfo  {journal}
  {J.~Chem.\ Phys.}\ }\textbf {\bibinfo {volume} {122}},\ \bibinfo {pages}
  {114507} (\bibinfo {year} {2005})}\BibitemShut {NoStop}%
\bibitem [{\citenamefont {Handel}\ \emph {et~al.}(2008)\citenamefont {Handel},
  \citenamefont {Davidchack}, \citenamefont {Anwar},\ and\ \citenamefont
  {Brukhno}}]{Handel2008}%
  \BibitemOpen
  \bibfield  {author} {\bibinfo {author} {\bibfnamefont {R.}~\bibnamefont
  {Handel}}, \bibinfo {author} {\bibfnamefont {R.~L.}\ \bibnamefont
  {Davidchack}}, \bibinfo {author} {\bibfnamefont {J.}~\bibnamefont {Anwar}}, \
  and\ \bibinfo {author} {\bibfnamefont {A.}~\bibnamefont {Brukhno}},\ }\href
  {\doibase 10.1103/PhysRevLett.100.036104} {\bibfield  {journal} {\bibinfo
  {journal} {Phys.\ Rev.\ Lett.}\ }\textbf {\bibinfo {volume} {100}},\ \bibinfo
  {pages} {036104} (\bibinfo {year} {2008})}\BibitemShut {NoStop}%
\bibitem [{\citenamefont {Bigg}(1953)}]{Bigg1953}%
  \BibitemOpen
  \bibfield  {author} {\bibinfo {author} {\bibfnamefont {E.~K.}\ \bibnamefont
  {Bigg}},\ }\href {\doibase 10.1088/0370-1301/66/8/309} {\bibfield  {journal}
  {\bibinfo  {journal} {Proc.~Phys.~Soc.~B}\ }\textbf {\bibinfo {volume}
  {66}},\ \bibinfo {pages} {688} (\bibinfo {year} {1953})}\BibitemShut
  {NoStop}%
\bibitem [{\citenamefont {Wood}\ and\ \citenamefont {Walton}(1970)}]{Wood1970}%
  \BibitemOpen
  \bibfield  {author} {\bibinfo {author} {\bibfnamefont {G.~R.}\ \bibnamefont
  {Wood}}\ and\ \bibinfo {author} {\bibfnamefont {A.~G.}\ \bibnamefont
  {Walton}},\ }\href {\doibase 10.1063/1.1659359} {\bibfield  {journal}
  {\bibinfo  {journal} {J.~Appl.\ Phys.}\ }\textbf {\bibinfo {volume} {41}},\
  \bibinfo {pages} {3027} (\bibinfo {year} {1970})}\BibitemShut {NoStop}%
\bibitem [{\citenamefont {Molinero}\ and\ \citenamefont
  {Moore}(2009)}]{Molinero2009}%
  \BibitemOpen
  \bibfield  {author} {\bibinfo {author} {\bibfnamefont {V.}~\bibnamefont
  {Molinero}}\ and\ \bibinfo {author} {\bibfnamefont {E.~B.}\ \bibnamefont
  {Moore}},\ }\href {\doibase 10.1021/jp805227c} {\bibfield  {journal}
  {\bibinfo  {journal} {J.~Phys.\ Chem.~B}\ }\textbf {\bibinfo {volume}
  {113}},\ \bibinfo {pages} {4008} (\bibinfo {year} {2009})}\BibitemShut
  {NoStop}%
\bibitem [{\citenamefont {Moore}\ and\ \citenamefont
  {Molinero}(2009)}]{Moore2009}%
  \BibitemOpen
  \bibfield  {author} {\bibinfo {author} {\bibfnamefont {E.~B.}\ \bibnamefont
  {Moore}}\ and\ \bibinfo {author} {\bibfnamefont {V.}~\bibnamefont
  {Molinero}},\ }\href {\doibase 10.1063/1.3158470} {\bibfield  {journal}
  {\bibinfo  {journal} {J.~Chem.\ Phys.}\ }\textbf {\bibinfo {volume} {130}},\
  \bibinfo {pages} {244505} (\bibinfo {year} {2009})}\BibitemShut {NoStop}%
\bibitem [{\citenamefont {Limmer}\ and\ \citenamefont
  {Chandler}(2011)}]{Limmer2011}%
  \BibitemOpen
  \bibfield  {author} {\bibinfo {author} {\bibfnamefont {D.~T.}\ \bibnamefont
  {Limmer}}\ and\ \bibinfo {author} {\bibfnamefont {D.}~\bibnamefont
  {Chandler}},\ }\href {\doibase 10.1063/1.3643333} {\bibfield  {journal}
  {\bibinfo  {journal} {J.~Chem.\ Phys.}\ }\textbf {\bibinfo {volume} {135}},\
  \bibinfo {pages} {134503} (\bibinfo {year} {2011})}\BibitemShut {NoStop}%
\bibitem [{\citenamefont {Moore}\ and\ \citenamefont
  {Molinero}(2010)}]{Moore2010}%
  \BibitemOpen
  \bibfield  {author} {\bibinfo {author} {\bibfnamefont {E.~B.}\ \bibnamefont
  {Moore}}\ and\ \bibinfo {author} {\bibfnamefont {V.}~\bibnamefont
  {Molinero}},\ }\href {\doibase 10.1063/1.3451112} {\bibfield  {journal}
  {\bibinfo  {journal} {J.~Chem.\ Phys.}\ }\textbf {\bibinfo {volume} {132}},\
  \bibinfo {pages} {244504} (\bibinfo {year} {2010})}\BibitemShut {NoStop}%
\bibitem [{\citenamefont {Moore}\ and\ \citenamefont
  {Molinero}(2011{\natexlab{a}})}]{Moore2011}%
  \BibitemOpen
  \bibfield  {author} {\bibinfo {author} {\bibfnamefont {E.~B.}\ \bibnamefont
  {Moore}}\ and\ \bibinfo {author} {\bibfnamefont {V.}~\bibnamefont
  {Molinero}},\ }\href {\doibase 10.1038/nature10586} {\bibfield  {journal}
  {\bibinfo  {journal} {Nature}\ }\textbf {\bibinfo {volume} {479}},\ \bibinfo
  {pages} {506} (\bibinfo {year} {2011}{\natexlab{a}})}\BibitemShut {NoStop}%
\bibitem [{\citenamefont {Li}\ \emph {et~al.}(2011)\citenamefont {Li},
  \citenamefont {Donadio}, \citenamefont {Russo},\ and\ \citenamefont
  {Galli}}]{Li2011}%
  \BibitemOpen
  \bibfield  {author} {\bibinfo {author} {\bibfnamefont {T.}~\bibnamefont
  {Li}}, \bibinfo {author} {\bibfnamefont {D.}~\bibnamefont {Donadio}},
  \bibinfo {author} {\bibfnamefont {G.}~\bibnamefont {Russo}}, \ and\ \bibinfo
  {author} {\bibfnamefont {G.}~\bibnamefont {Galli}},\ }\href {\doibase
  10.1039/C1CP22167A} {\bibfield  {journal} {\bibinfo  {journal} {Phys.\ Chem.\
  Chem.\ Phys.}\ }\textbf {\bibinfo {volume} {13}},\ \bibinfo {pages} {19807}
  (\bibinfo {year} {2011})}\BibitemShut {NoStop}%
\bibitem [{\citenamefont {Kastelowitz}, \citenamefont {Johnston},\ and\
  \citenamefont {Molinero}(2010)}]{Kastelowitz2010}%
  \BibitemOpen
  \bibfield  {author} {\bibinfo {author} {\bibfnamefont {N.}~\bibnamefont
  {Kastelowitz}}, \bibinfo {author} {\bibfnamefont {J.~C.}\ \bibnamefont
  {Johnston}}, \ and\ \bibinfo {author} {\bibfnamefont {V.}~\bibnamefont
  {Molinero}},\ }\href {\doibase 10.1063/1.3368793} {\bibfield  {journal}
  {\bibinfo  {journal} {J.~Chem.\ Phys.}\ }\textbf {\bibinfo {volume} {132}},\
  \bibinfo {pages} {124511} (\bibinfo {year} {2010})}\BibitemShut {NoStop}%
\bibitem [{\citenamefont {Moore}\ \emph {et~al.}(2010)\citenamefont {Moore},
  \citenamefont {de~la Llave}, \citenamefont {Welke}, \citenamefont
  {Scherlis},\ and\ \citenamefont {Molinero}}]{Moore2010b}%
  \BibitemOpen
  \bibfield  {author} {\bibinfo {author} {\bibfnamefont {E.~B.}\ \bibnamefont
  {Moore}}, \bibinfo {author} {\bibfnamefont {E.}~\bibnamefont {de~la Llave}},
  \bibinfo {author} {\bibfnamefont {K.}~\bibnamefont {Welke}}, \bibinfo
  {author} {\bibfnamefont {D.~A.}\ \bibnamefont {Scherlis}}, \ and\ \bibinfo
  {author} {\bibfnamefont {V.}~\bibnamefont {Molinero}},\ }\href {\doibase
  10.1039/B919724A} {\bibfield  {journal} {\bibinfo  {journal} {Phys.\ Chem.\
  Chem.\ Phys.}\ }\textbf {\bibinfo {volume} {12}},\ \bibinfo {pages} {4124}
  (\bibinfo {year} {2010})}\BibitemShut {NoStop}%
\bibitem [{\citenamefont {Stillinger}\ and\ \citenamefont
  {Weber}(1985)}]{Stillinger1985}%
  \BibitemOpen
  \bibfield  {author} {\bibinfo {author} {\bibfnamefont {F.~H.}\ \bibnamefont
  {Stillinger}}\ and\ \bibinfo {author} {\bibfnamefont {T.~A.}\ \bibnamefont
  {Weber}},\ }\href {\doibase 10.1103/PhysRevB.31.5262} {\bibfield  {journal}
  {\bibinfo  {journal} {Phys.\ Rev.\ B}\ }\textbf {\bibinfo {volume} {31}},\
  \bibinfo {pages} {5262} (\bibinfo {year} {1985})}\BibitemShut {NoStop}%
\bibitem [{\citenamefont {Torrie}\ and\ \citenamefont
  {Valleau}(1977)}]{Torrie1977}%
  \BibitemOpen
  \bibfield  {author} {\bibinfo {author} {\bibfnamefont {G.~M.}\ \bibnamefont
  {Torrie}}\ and\ \bibinfo {author} {\bibfnamefont {J.~P.}\ \bibnamefont
  {Valleau}},\ }\href {\doibase 10.1016/0021-9991(77)90121-8} {\bibfield
  {journal} {\bibinfo  {journal} {J.~Comput.~Phys.}\ }\textbf {\bibinfo
  {volume} {23}},\ \bibinfo {pages} {187} (\bibinfo {year} {1977})}\BibitemShut
  {NoStop}%
\bibitem [{\citenamefont {Chandler}(1987)}]{Chandler1987}%
  \BibitemOpen
  \bibfield  {author} {\bibinfo {author} {\bibfnamefont {D.}~\bibnamefont
  {Chandler}},\ }\href@noop {} {\emph {\bibinfo {title} {Introduction to modern
  statistical mechanics}}}\ (\bibinfo  {publisher} {Oxford University Press},\
  \bibinfo {address} {New York},\ \bibinfo {year} {1987})\BibitemShut {NoStop}%
\bibitem [{\citenamefont {Romano}, \citenamefont {Sanz},\ and\ \citenamefont
  {Sciortino}(2011)}]{Romano2011}%
  \BibitemOpen
  \bibfield  {author} {\bibinfo {author} {\bibfnamefont {F.}~\bibnamefont
  {Romano}}, \bibinfo {author} {\bibfnamefont {E.}~\bibnamefont {Sanz}}, \ and\
  \bibinfo {author} {\bibfnamefont {F.}~\bibnamefont {Sciortino}},\ }\href
  {\doibase 10.1063/1.3578182} {\bibfield  {journal} {\bibinfo  {journal}
  {J.~Chem.\ Phys.}\ }\textbf {\bibinfo {volume} {134}},\ \bibinfo {pages}
  {174502} (\bibinfo {year} {2011})}\BibitemShut {NoStop}%
\bibitem [{\citenamefont {Blanco}, \citenamefont {Fl\'{o}rez},\ and\
  \citenamefont {Bermejo}(1997)}]{Blanco1997}%
  \BibitemOpen
  \bibfield  {author} {\bibinfo {author} {\bibfnamefont {M.~A.}\ \bibnamefont
  {Blanco}}, \bibinfo {author} {\bibfnamefont {M.}~\bibnamefont {Fl\'{o}rez}},
  \ and\ \bibinfo {author} {\bibfnamefont {M.}~\bibnamefont {Bermejo}},\ }\href
  {\doibase 10.1016/S0166-1280(97)00185-1} {\bibfield  {journal} {\bibinfo
  {journal} {J.\ Mol.\ Struc.--Theochem.}\ }\textbf {\bibinfo {volume} {419}},\
  \bibinfo {pages} {19} (\bibinfo {year} {1997})}\BibitemShut {NoStop}%
\bibitem [{\citenamefont {Plimpton}(1995)}]{Plimpton1995}%
  \BibitemOpen
  \bibfield  {author} {\bibinfo {author} {\bibfnamefont {S.}~\bibnamefont
  {Plimpton}},\ }\href {\doibase 10.1006/jcph.1995.1039} {\bibfield  {journal}
  {\bibinfo  {journal} {J.~Comp.~Phys.}\ }\textbf {\bibinfo {volume} {117}},\
  \bibinfo {pages} {1} (\bibinfo {year} {1995})}\BibitemShut {NoStop}%
\bibitem [{\citenamefont {Auer}\ and\ \citenamefont
  {Frenkel}(2001)}]{Auer2001}%
  \BibitemOpen
  \bibfield  {author} {\bibinfo {author} {\bibfnamefont {S.}~\bibnamefont
  {Auer}}\ and\ \bibinfo {author} {\bibfnamefont {D.}~\bibnamefont {Frenkel}},\
  }\href {\doibase 10.1038/35059035} {\bibfield  {journal} {\bibinfo  {journal}
  {Nature}\ }\textbf {\bibinfo {volume} {409}},\ \bibinfo {pages} {1020}
  (\bibinfo {year} {2001})}\BibitemShut {NoStop}%
\bibitem [{\citenamefont {Auer}\ and\ \citenamefont
  {Frenkel}(2002)}]{Auer2002}%
  \BibitemOpen
  \bibfield  {author} {\bibinfo {author} {\bibfnamefont {S.}~\bibnamefont
  {Auer}}\ and\ \bibinfo {author} {\bibfnamefont {D.}~\bibnamefont {Frenkel}},\
  }\href {\doibase 10.1088/0953-8984/14/33/308} {\bibfield  {journal} {\bibinfo
   {journal} {J.\ Phys.:\ Cond.\ Matt.}\ }\textbf {\bibinfo {volume} {14}},\
  \bibinfo {pages} {7667} (\bibinfo {year} {2002})}\BibitemShut {NoStop}%
\bibitem [{\citenamefont {Filion}\ \emph {et~al.}(2010)\citenamefont {Filion},
  \citenamefont {Hermes}, \citenamefont {Ni},\ and\ \citenamefont
  {Dijkstra}}]{Filion2010}%
  \BibitemOpen
  \bibfield  {author} {\bibinfo {author} {\bibfnamefont {L.}~\bibnamefont
  {Filion}}, \bibinfo {author} {\bibfnamefont {M.}~\bibnamefont {Hermes}},
  \bibinfo {author} {\bibfnamefont {R.}~\bibnamefont {Ni}}, \ and\ \bibinfo
  {author} {\bibfnamefont {M.}~\bibnamefont {Dijkstra}},\ }\href {\doibase
  10.1063/1.3506838} {\bibfield  {journal} {\bibinfo  {journal} {J.~Chem.\
  Phys.}\ }\textbf {\bibinfo {volume} {133}},\ \bibinfo {pages} {244115}
  (\bibinfo {year} {2010})}\BibitemShut {NoStop}%
\bibitem [{\citenamefont {Chandler}(1978)}]{Chandler1978}%
  \BibitemOpen
  \bibfield  {author} {\bibinfo {author} {\bibfnamefont {D.}~\bibnamefont
  {Chandler}},\ }\href {\doibase 10.1063/1.436049} {\bibfield  {journal}
  {\bibinfo  {journal} {J.~Chem.\ Phys.}\ }\textbf {\bibinfo {volume} {68}},\
  \bibinfo {pages} {2959} (\bibinfo {year} {1978})}\BibitemShut {NoStop}%
\bibitem [{\citenamefont {Ruiz-Montero}, \citenamefont {Frenkel},\ and\
  \citenamefont {Brey}(1997)}]{RuizMontero1997}%
  \BibitemOpen
  \bibfield  {author} {\bibinfo {author} {\bibfnamefont {M.~J.}\ \bibnamefont
  {Ruiz-Montero}}, \bibinfo {author} {\bibfnamefont {D.}~\bibnamefont
  {Frenkel}}, \ and\ \bibinfo {author} {\bibfnamefont {J.~J.}\ \bibnamefont
  {Brey}},\ }\href {\doibase 10.1080/002689797171922} {\bibfield  {journal}
  {\bibinfo  {journal} {Mol.\ Phys.}\ }\textbf {\bibinfo {volume} {90}},\
  \bibinfo {pages} {925} (\bibinfo {year} {1997})}\BibitemShut {NoStop}%
\bibitem [{Note4()}]{Note4}%
  \BibitemOpen
  \bibinfo {note} {{\protect \color {red}The five seed clusters chosen for each
  system comprised approximately 35 particles. In the first umbrella sampling
  window, we allowed the cluster to melt completely and then to regrow within
  the window. The point $\upDelta G=0$ corresponds to the supercooled liquid
  state for all simulations. Matching the free energy difference between the
  first and the second window is somewhat problematic, as the equilibrated
  state in the first window is, given that the clusters are allowed to melt
  completely, the same as in the simulations grown from the supercooled liquid
  directly. However, we suggest the free energy difference between the
  simulations with different starting points arises at a later stage, when the
  clusters cease to be interconvertible. We justify this assertion by
  performing umbrella sampling simulations on an intermediate window between
  the first and the second one, in which the cluster is not allowed to melt
  completely; we observe no difference between the free energies reported in
  Fig.~\ref {fig-MW-energybarrier} and in this overlapping window, which
  suggests that, although the resulting ice nuclei are somewhat different
  between the equilibrated states in the overlapping regions of the first two
  umbrella sampling windows in seeded simulations, they have the same free
  energy.}}\BibitemShut {Stop}%
\bibitem [{Note5()}]{Note5}%
  \BibitemOpen
  \bibinfo {note} {For example, if an eclipsed bond is with a particle
  classified as being liquid, it is unclear whether this should qualify a
  particle as being hexagonal.}\BibitemShut {Stop}%
\bibitem [{Note6()}]{Note6}%
  \BibitemOpen
  \bibinfo {note} {Core ice particles are ice particles in the largest
  crystalline cluster whose four neighbours (particles within \SI
  {3.6}{\angstrom }) all also belong to the cluster; such particles are
  classified as being hexagonal if they have one eclipsed and three staggered
  connections to neighbours.}\BibitemShut {Stop}%
\bibitem [{\citenamefont {Lechner}, \citenamefont {Dellago},\ and\
  \citenamefont {Bolhuis}(2011)}]{Lechner2011}%
  \BibitemOpen
  \bibfield  {author} {\bibinfo {author} {\bibfnamefont {W.}~\bibnamefont
  {Lechner}}, \bibinfo {author} {\bibfnamefont {C.}~\bibnamefont {Dellago}}, \
  and\ \bibinfo {author} {\bibfnamefont {P.~G.}\ \bibnamefont {Bolhuis}},\
  }\href {\doibase 10.1103/PhysRevLett.106.085701} {\bibfield  {journal}
  {\bibinfo  {journal} {Phys.\ Rev.\ Lett.}\ }\textbf {\bibinfo {volume}
  {106}},\ \bibinfo {pages} {085701} (\bibinfo {year} {2011})}\BibitemShut
  {NoStop}%
\bibitem [{\citenamefont {Moore}\ and\ \citenamefont
  {Molinero}(2011{\natexlab{b}})}]{Moore2011b}%
  \BibitemOpen
  \bibfield  {author} {\bibinfo {author} {\bibfnamefont {E.~B.}\ \bibnamefont
  {Moore}}\ and\ \bibinfo {author} {\bibfnamefont {V.}~\bibnamefont
  {Molinero}},\ }\href {\doibase 10.1039/C1CP22022E} {\bibfield  {journal}
  {\bibinfo  {journal} {Phys.\ Chem.\ Chem.\ Phys.}\ }\textbf {\bibinfo
  {volume} {13}},\ \bibinfo {pages} {20008} (\bibinfo {year}
  {2011}{\natexlab{b}})}\BibitemShut {NoStop}%
\bibitem [{\citenamefont {Sanz}\ \emph {et~al.}(2007)\citenamefont {Sanz},
  \citenamefont {Valeriani}, \citenamefont {Frenkel},\ and\ \citenamefont
  {Dijkstra}}]{Sanz2007}%
  \BibitemOpen
  \bibfield  {author} {\bibinfo {author} {\bibfnamefont {E.}~\bibnamefont
  {Sanz}}, \bibinfo {author} {\bibfnamefont {C.}~\bibnamefont {Valeriani}},
  \bibinfo {author} {\bibfnamefont {D.}~\bibnamefont {Frenkel}}, \ and\
  \bibinfo {author} {\bibfnamefont {M.}~\bibnamefont {Dijkstra}},\ }\href
  {\doibase 10.1103/PhysRevLett.99.055501} {\bibfield  {journal} {\bibinfo
  {journal} {Phys.\ Rev.\ Lett.}\ }\textbf {\bibinfo {volume} {99}},\ \bibinfo
  {pages} {055501} (\bibinfo {year} {2007})}\BibitemShut {NoStop}%
\bibitem [{Note7()}]{Note7}%
  \BibitemOpen
  \bibinfo {note} {We assume that $t_2$ corresponds directly to $\gamma
  (36\uppi /\rho ^2)^{1/3}$; the remaining two fitted terms serve primarily to
  shift the curve appropriately, as the simulation data curve goes through zero
  at a cluster size of 2; this is a consequence of the classification parameter
  not being well-suited to extremely small clusters.}\BibitemShut {Stop}%
\bibitem [{\citenamefont {Beaucage}\ and\ \citenamefont
  {Mousseau}(2005)}]{Beaucage2005}%
  \BibitemOpen
  \bibfield  {author} {\bibinfo {author} {\bibfnamefont {P.}~\bibnamefont
  {Beaucage}}\ and\ \bibinfo {author} {\bibfnamefont {N.}~\bibnamefont
  {Mousseau}},\ }\href {\doibase 10.1103/PhysRevB.71.094102} {\bibfield
  {journal} {\bibinfo  {journal} {Phys.\ Rev.\ B}\ }\textbf {\bibinfo {volume}
  {71}},\ \bibinfo {pages} {094102} (\bibinfo {year} {2005})}\BibitemShut
  {NoStop}%
\bibitem [{\citenamefont {Rudnick}\ and\ \citenamefont
  {Gaspari}(1986)}]{Rudnick1986}%
  \BibitemOpen
  \bibfield  {author} {\bibinfo {author} {\bibfnamefont {J.}~\bibnamefont
  {Rudnick}}\ and\ \bibinfo {author} {\bibfnamefont {G.}~\bibnamefont
  {Gaspari}},\ }\href {\doibase 10.1088/0305-4470/19/4/004} {\bibfield
  {journal} {\bibinfo  {journal} {J.~Phys.~A:~Math.~Gen.}\ }\textbf {\bibinfo
  {volume} {19}},\ \bibinfo {pages} {L191} (\bibinfo {year}
  {1986})}\BibitemShut {NoStop}%
\bibitem [{\citenamefont {Chau}\ and\ \citenamefont
  {Hardwick}(1998)}]{Chau1998}%
  \BibitemOpen
  \bibfield  {author} {\bibinfo {author} {\bibfnamefont {P.-L.}\ \bibnamefont
  {Chau}}\ and\ \bibinfo {author} {\bibfnamefont {A.~J.}\ \bibnamefont
  {Hardwick}},\ }\href {\doibase 10.1080/002689798169195} {\bibfield  {journal}
  {\bibinfo  {journal} {Mol.\ Phys.}\ }\textbf {\bibinfo {volume} {93}},\
  \bibinfo {pages} {511} (\bibinfo {year} {1998})}\BibitemShut {NoStop}%
\bibitem [{\citenamefont {Errington}\ and\ \citenamefont
  {Debenedetti}(2001)}]{Errington2001}%
  \BibitemOpen
  \bibfield  {author} {\bibinfo {author} {\bibfnamefont {J.~R.}\ \bibnamefont
  {Errington}}\ and\ \bibinfo {author} {\bibfnamefont {P.~G.}\ \bibnamefont
  {Debenedetti}},\ }\href {\doibase 10.1038/35053024} {\bibfield  {journal}
  {\bibinfo  {journal} {Nature}\ }\textbf {\bibinfo {volume} {409}},\ \bibinfo
  {pages} {318} (\bibinfo {year} {2001})}\BibitemShut {NoStop}%
\bibitem [{Note8()}]{Note8}%
  \BibitemOpen
  \bibinfo {note} {We remark here that Quigley and Rodger suggest that $\zeta $
  tends to unity for perfect tetrahedral networks;\cite {Quigley2009} however,
  since the tetrahedral angle is $\protect \qopname \relax o{arccos}(-1/3)$,
  the sum in Eq.~\ref {eqn-ChauHardwick} will clearly tend to zero when the
  network is tetrahedral. We therefore assume that Quigley and Rodger actually
  reported their results as a function of $1-\zeta $.}\BibitemShut {Stop}%
\bibitem [{\citenamefont {Quigley}\ and\ \citenamefont
  {Rodger}(2009)}]{Quigley2009}%
  \BibitemOpen
  \bibfield  {author} {\bibinfo {author} {\bibfnamefont {D.}~\bibnamefont
  {Quigley}}\ and\ \bibinfo {author} {\bibfnamefont {P.~M.}\ \bibnamefont
  {Rodger}},\ }\href {\doibase 10.1080/08927020802647280} {\bibfield  {journal}
  {\bibinfo  {journal} {Mol.\ Simulat.}\ }\textbf {\bibinfo {volume} {35}},\
  \bibinfo {pages} {613} (\bibinfo {year} {2009})}\BibitemShut {NoStop}%
\bibitem [{\citenamefont {Leyssale}, \citenamefont {Delhommelle},\ and\
  \citenamefont {Millot}(2005)}]{Leyssale2005}%
  \BibitemOpen
  \bibfield  {author} {\bibinfo {author} {\bibfnamefont {J.-M.}\ \bibnamefont
  {Leyssale}}, \bibinfo {author} {\bibfnamefont {J.}~\bibnamefont
  {Delhommelle}}, \ and\ \bibinfo {author} {\bibfnamefont {C.}~\bibnamefont
  {Millot}},\ }\href {\doibase 10.1063/1.1862626} {\bibfield  {journal}
  {\bibinfo  {journal} {J.~Chem.\ Phys.}\ }\textbf {\bibinfo {volume} {122}},\
  \bibinfo {pages} {104510} (\bibinfo {year} {2005})}\BibitemShut {NoStop}%
\bibitem [{\citenamefont {Allen}, \citenamefont {Warren},\ and\ \citenamefont
  {ten Wolde}(2005)}]{Allen2005}%
  \BibitemOpen
  \bibfield  {author} {\bibinfo {author} {\bibfnamefont {R.~J.}\ \bibnamefont
  {Allen}}, \bibinfo {author} {\bibfnamefont {P.~B.}\ \bibnamefont {Warren}}, \
  and\ \bibinfo {author} {\bibfnamefont {P.~R.}\ \bibnamefont {ten Wolde}},\
  }\href {\doibase 10.1103/PhysRevLett.94.018104} {\bibfield  {journal}
  {\bibinfo  {journal} {Phys.\ Rev.\ Lett.}\ }\textbf {\bibinfo {volume}
  {94}},\ \bibinfo {pages} {018104} (\bibinfo {year} {2005})}\BibitemShut
  {NoStop}%
\bibitem [{\citenamefont {Allen}, \citenamefont {Frenkel},\ and\ \citenamefont
  {ten Wolde}(2006{\natexlab{a}})}]{Allen2006}%
  \BibitemOpen
  \bibfield  {author} {\bibinfo {author} {\bibfnamefont {R.~J.}\ \bibnamefont
  {Allen}}, \bibinfo {author} {\bibfnamefont {D.}~\bibnamefont {Frenkel}}, \
  and\ \bibinfo {author} {\bibfnamefont {P.~R.}\ \bibnamefont {ten Wolde}},\
  }\href {\doibase 10.1063/1.2198827} {\bibfield  {journal} {\bibinfo
  {journal} {J.~Chem.\ Phys.}\ }\textbf {\bibinfo {volume} {124}},\ \bibinfo
  {pages} {194111} (\bibinfo {year} {2006}{\natexlab{a}})}\BibitemShut
  {NoStop}%
\bibitem [{\citenamefont {Allen}, \citenamefont {Frenkel},\ and\ \citenamefont
  {ten Wolde}(2006{\natexlab{b}})}]{Allen2006b}%
  \BibitemOpen
  \bibfield  {author} {\bibinfo {author} {\bibfnamefont {R.~J.}\ \bibnamefont
  {Allen}}, \bibinfo {author} {\bibfnamefont {D.}~\bibnamefont {Frenkel}}, \
  and\ \bibinfo {author} {\bibfnamefont {P.~R.}\ \bibnamefont {ten Wolde}},\
  }\href {\doibase 10.1063/1.2140273} {\bibfield  {journal} {\bibinfo
  {journal} {J.~Chem.\ Phys.}\ }\textbf {\bibinfo {volume} {124}},\ \bibinfo
  {pages} {024102} (\bibinfo {year} {2006}{\natexlab{b}})}\BibitemShut
  {NoStop}%
\bibitem [{\citenamefont {Wang}, \citenamefont {Valeriani},\ and\ \citenamefont
  {Frenkel}(2009)}]{Wang2009}%
  \BibitemOpen
  \bibfield  {author} {\bibinfo {author} {\bibfnamefont {Z.-J.}\ \bibnamefont
  {Wang}}, \bibinfo {author} {\bibfnamefont {C.}~\bibnamefont {Valeriani}}, \
  and\ \bibinfo {author} {\bibfnamefont {D.}~\bibnamefont {Frenkel}},\ }\href
  {\doibase 10.1021/jp807727p} {\bibfield  {journal} {\bibinfo  {journal}
  {J.~Phys.\ Chem.~B}\ }\textbf {\bibinfo {volume} {113}},\ \bibinfo {pages}
  {3776} (\bibinfo {year} {2009})}\BibitemShut {NoStop}%
\bibitem [{\citenamefont {Li}, \citenamefont {Donadio},\ and\ \citenamefont
  {Galli}(2009)}]{Li2009b}%
  \BibitemOpen
  \bibfield  {author} {\bibinfo {author} {\bibfnamefont {T.}~\bibnamefont
  {Li}}, \bibinfo {author} {\bibfnamefont {D.}~\bibnamefont {Donadio}}, \ and\
  \bibinfo {author} {\bibfnamefont {G.}~\bibnamefont {Galli}},\ }\href
  {\doibase 10.1063/1.3268346} {\bibfield  {journal} {\bibinfo  {journal}
  {J.~Chem.\ Phys.}\ }\textbf {\bibinfo {volume} {131}},\ \bibinfo {pages}
  {224519} (\bibinfo {year} {2009})}\BibitemShut {NoStop}%
\end{thebibliography}%
\end{document}